\documentclass[12pt]{article}
\usepackage[margin=1in]{geometry} 
\usepackage{amssymb} 
\usepackage{amsmath} 
\usepackage{amsthm} 
\usepackage{graphicx} 
\usepackage{booktabs} 
\usepackage{natbib} 
\usepackage{pgfplots} 
\usepgfplotslibrary{fillbetween} 
\usepackage{bm} 
\usepackage{xcolor} 
\definecolor{royalblue}{HTML}{0000CD} 
\usepackage[colorlinks=true]{hyperref} 
\hypersetup{
    colorlinks=true,
    allcolors=royalblue,
} 
\usepackage{threeparttable} 
\usepackage{thmtools} 
\allowdisplaybreaks 
\usepackage{mathtools} 

\theoremstyle{definition}

\providecommand{\keywords}[1]
{
  \noindent\small	
  \textbf{\textit{Keywords:}} #1
}

\usepackage{caption} 
\usepackage{subcaption} 
\usepackage{hyperref}
\usepackage{setspace} 
\usepackage{enumitem} 
\usepackage{array}
\usepackage{placeins} 
\usepackage{siunitx} 
\newcolumntype{L}{>{\centering\arraybackslash}m{3cm}} 

\usepackage{rotating}
\usepackage{soul}

\begin{document}

\title{Clustering
Plasma Concentration-Time Curves: Applications of Unsupervised Learning in Pharmacogenomics}

\author{
  Jackson P. Lautier\footnote{Department of Mathematical Sciences,
  Bentley University, Waltham, MA, USA}
  \and
  Stella Grosser\footnote{Office of Biostatistics, Center for Drug Evaluation
  and Research, U.S. Food and Drug Administration, Silver Spring, MD, USA}
  \and
  Jessica Kim\footnotemark[2]
  \and
  Hyewon Kim\footnote{Office of Clinical Pharmacology, Center for Drug
  Evaluation and Research, U.S. Food and Drug Administration, Silver Spring,
  MD, USA}
  \and
  Junghi Kim\footnotemark[2]
  \thanks{Corresponding to
  junghi.kim@fda.hhs.gov, 10903 New Hampshire Avenue, Silver Spring, MD, USA}
}

\date{}

\maketitle

\begin{abstract}
Pharmaceutical researchers are continually searching for techniques to improve
both drug development processes and patient outcomes. An area of recent
interest is the potential for machine learning (ML) applications within
pharmacology.  One such application not yet given close study is the
unsupervised clustering of plasma concentration-time curves, hereafter,
pharmacokinetic (PK) curves.
In this paper, we present our findings on how to cluster PK curves by their similarity.  Specifically, we find clustering to be effective
at identifying similar-shaped PK curves and informative for
understanding patterns within each cluster of PK curves.
Because PK curves are time series data objects,
our approach utilizes the extensive body of research related to the
clustering of time series data as a starting point.  As such, we examine many dissimilarity measures between time series data objects to
find those most suitable for PK curves.  We
identify Euclidean distance as generally most appropriate for clustering PK
curves, and we further show that dynamic time warping, Fr\'{e}chet, and
structure-based measures of dissimilarity like correlation may produce
unexpected results.
As an illustration, we apply these methods in a case study with 250 PK
curves used in a previous pharmacogenomic study.
Our case study finds that an unsupervised ML clustering with
Euclidean distance, without any subject genetic information,
is able to independently validate the same conclusions as the reference
pharmacogenomic results.
To our knowledge, this is the first such demonstration.
Further, the case study demonstrates how the clustering of PK curves may
generate insights that could be difficult to perceive solely with
population level summary statistics of PK metrics.

\bigskip


\end{abstract}

\doublespacing

\section{Introduction}
\label{sec:intro}

Plasma concentration time-curves or pharmacokinetic (PK) curves are
generated by plotting drug concentration levels in plasma samples at various
time intervals after the administration of a drug product
\citep{shargel_2016}.  As such, they are an essential component of
characterizing drug disposition, which is an important prerequisite to
determine or modify dosing regimens for individuals and groups of patients
\citep{shargel_2016}.
\textcolor{black}{
Because of recent machine learning (ML) and artifical intelligence (AI)
work related generally to drug development \citep{zhang_2022} and patient
outcomes \citep{kumar_2022}, it is natural to suppose that any ML and AI
applications with the potential to enhance the understanding or interpretation
of PK curves would be of significant interest within the broader field
of pharmaceutical research.  Indeed, this is the case
\citep[e.g.,][]{koch_2020, zame_2020, mccomb_2021}.
Absent from these studies is the use of clustering techniques on data sets
of PK curves, however, and we thus present to our knowledge the first
applied study of clustering techniques for use on PK data.
}

\begin{figure}[t!]
    \centering
    \includegraphics[width=0.6\textwidth]{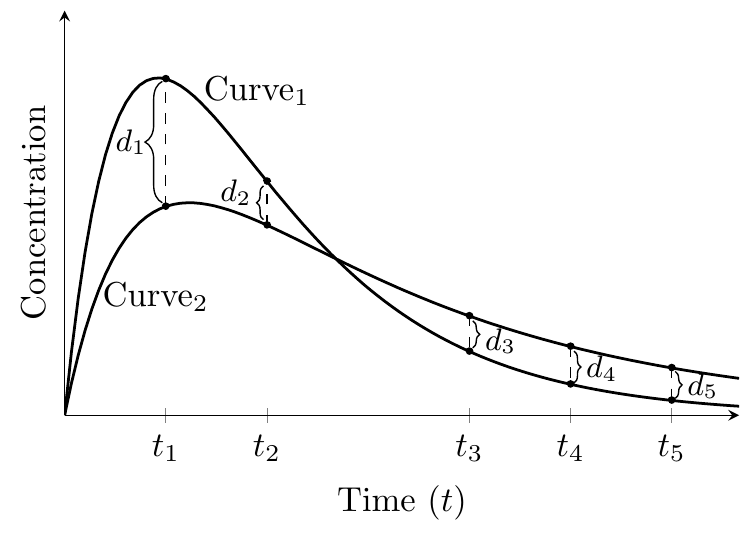}
    \caption{ {\bf Illustrative Example of Distance Between Two PK Curves}.
    {\scriptsize
    It is not straightforward to measure the dissimilarity between two
    PK curves for use within a ML clustering exercise.  The above is a
    geometric interpretation of ``distance'' between two PK curves with five
    plasma concentration sampling time points.}
    }   
    \label{fig:ex_pk_dist}
\end{figure}

\textcolor{black}{
We may use the observation that PK curves are time series data to
utilize existing literature related to the ML clustering of time series
data as a starting point to cluster similar PK curves. Broadly, time series
clustering is a technique for grouping time series data based on
their similarity.}
Clustering techniques for time series data have grown rapidly  and have been
applied successfully in a wide range of domains including medicine (e.g., personalized drug design), environmental science, and
many more \citep{javed_2020}.
In particular, there exists a large set of
well-established time series dissimilarity (or distance) measures \citep{montero_2014}.
It is not straightforward to define dissimilarity between PK curves,
and we illustrate a potential shape-based distance interpretation for
two PK curves with five concentration sampling points in
Figure~\ref{fig:ex_pk_dist}. The novelty of this paper is twofold. First,
we narrow down the broad field of potential
time series dissimilarity measures to select
a suitable one for use on PK curves.
Specifically, we find the following five dissimilarity
measures most applicable: correlation, Fr\'{e}chet, dynamic
time warping (DTW), temporal correlation coefficient (CORT),
and Euclidean.  Of these five, we
identify Euclidean distance as generally
most appropriate for clustering PK concentration curves, and we
summarize the pros and cons of all five of these dissimilarity measures
in Table~\ref{tab:dist_summary}.

Second, we present a novel case study to illustrate
the merits of clustering PK curves.  The case study objective is to use
ML clustering to independently analyze data from a concluded
pharmacogenomic study that attempted to tailor treatment strategies to
the individual patient level.
The case study data set consists of PK curves, along with genetic and
demographic information from 250 observations, and it spans nine Phase 1
studies.  As such, the concentration sampling time points vary between
PK curve data observations.  This is a well-known issue in time series
clustering \citep{montero_2014}.  While there are dissimilarity measures
in Section~\ref{sec:methods} that are able to handle PK curves with a
differing number of sampling times, we find such methods may not yield
desirable results (see Section~\ref{sec:methods} and\
the \hyperref[sec:diss_details]{Appendix} for details).
Hence, we find Euclidean distance using only the shared concentration
sampling time points to be most effective.  Lastly, the case study
demonstrates how clustering PK curves with Euclidean distance
can provide additional insights in comparison to PK
analysis based solely on PK parameters
(e.g.,
area-under-the-time-concentration curve (AUC),
maximum concentration ($C_{\max}$),
time-until-maximum concentration ($T_{\max}$), etc.).

The paper proceeds as follows. The methodological review occurs in
Section~\ref{sec:methods}.
The case study then follows in Section~\ref{sec:case_study}, and
Section~\ref{sec:disc} concludes.
For reference, the
\hyperref[sec:diss_details]{Appendix} provides extended details on the
dissimilarity measures of Section~\ref{sec:methods}, and the
Supplementary Material provides additional details on clustering
methods, cluster selection criteria, and the case study results.

\section{Methods}
\label{sec:methods}

Clustering algorithms are usually categorized by families, such as hierarchical
clustering, partition-based, model-based, and density-based clustering
\citep[e.g.,][]{hastie_2009, james_2013}.
We situate our paper within hierarchical clustering 
with a particular focus on similarity measures for PK curves and 
defer discussion of other algorithms to the Supplemental Material.
(Hierarchical clustering will also be used exclusively within the
case study, see Section~\ref{subsec:res}.)

An important step in hierarchical clustering is to find an appropriate distance or dissimilarity measure between data objects to be clustered, which is done in Section~\ref{subsec:diss_meas}.
Section~\ref{subsec:add_cons} then briefly reviews the well-known
problem of deciding on the final number of clusters
\citep[e.g.,][]{james_2013}.

\subsection{Hierarchical Clustering}
\label{subsec:hier_clst}

\begin{figure}[t!]
    \centering
    \includegraphics[width=\textwidth]{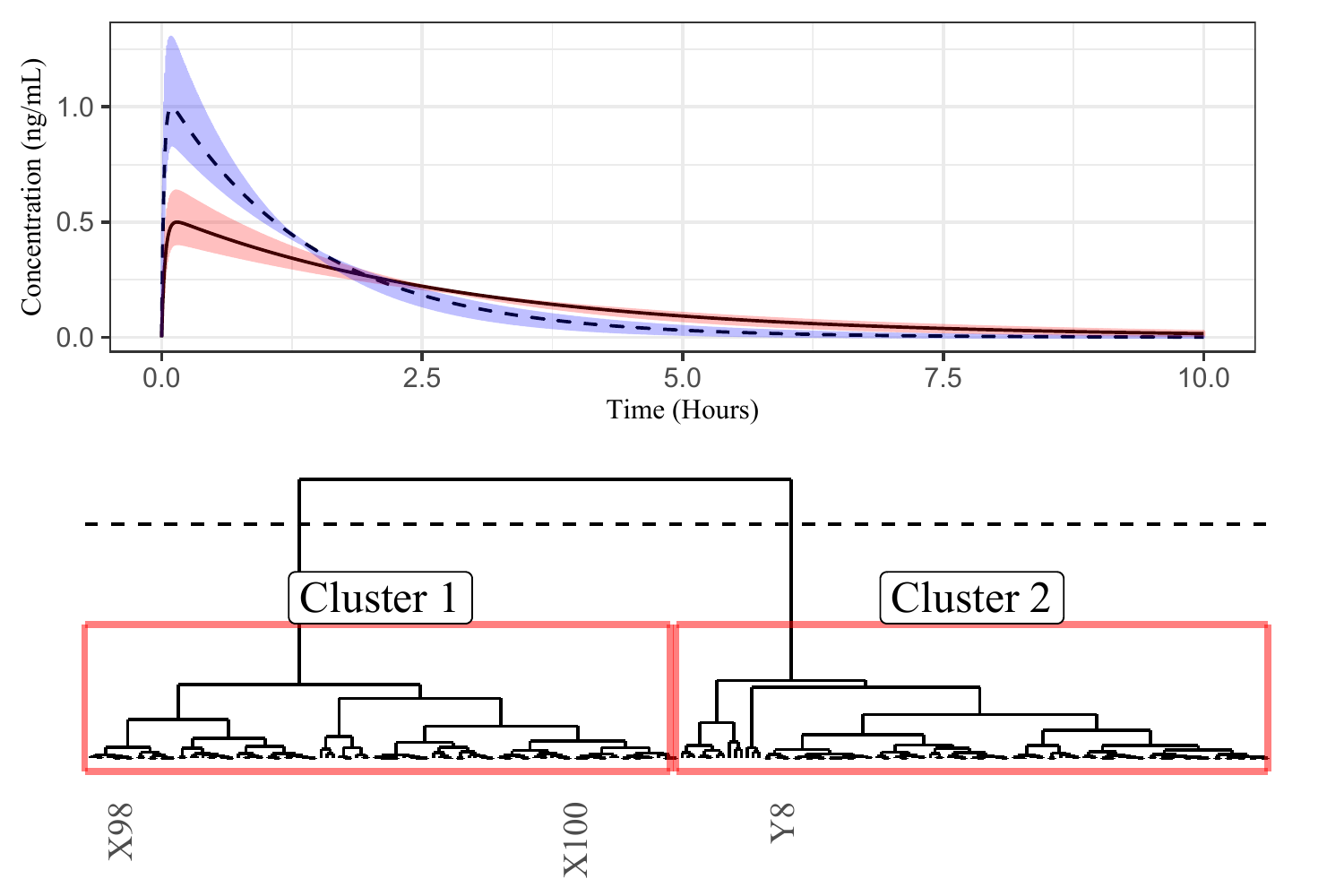}
    \caption{ {\bf An Illustration of Hierarchical Clustering for PK Curves}.
    {\scriptsize (Top)
    An example of two resulting groups of
    100 simulated PK curves with the group mean curves denoted
    by the dashed and solid black curves.  The simulated PK curves
    were generated using \citet{R_2022} with the package \texttt{linpk}
    \citep{rich_2022}. (Bottom) The red squares indicate the two resulting
    clusters using Euclidean distance \eqref{eq:euc_dist} for a given
    scenario. Simulated subjects X98, X100, and Y8 are labeled to demonstrate
    that it is a mistake to associate a proximity on the horizontal axis (i.e.,
    Y8 and X100) with similarity.  That is, subjects X98 and X100 fuse on the
    vertical axis much earlier than X100 and Y8 and are thus relatively more
    similar than X100 and Y8.  Indeed, even with two clusters selected, X100
    and Y8 are in separate clusters. The horizontal dashed line indicates
    where a ``cut'' to the dendrogram tree would occur to create two clusters
    (see Section~\ref{subsec:add_cons} for more details).
    }
    }
    \label{fig:hier_ill}
\end{figure}

Hierarchical clustering is a “bottom-up” clustering technique with an
attractive feature of producing a tree-based representation of observations
called a \textit{dendrogram}. Colloquially,
data objects that are relatively similar are to be grouped into the same
cluster, while data objects that are relatively dissimilar are to be grouped
into separate clusters. The dendrogram, therefore, represents the relationships
of similarity among all objects in a data set (see bottom of Figure~\ref{fig:hier_ill}).

At the beginning of the algorithm, each data object is treated as a single
cluster.
The two most similar clusters are then merged into a new single cluster, and
this new cluster becomes an updated data object with a value that is determined
by averaging its now two members.  There are other methods to determine the
new cluster value besides averaging, but we defer this discussion at present
for ease of exposition (the precise vernacular is \textit{linkage}, see
\citet{james_2013} for details).
The merging process continues until all original data objects
eventually merge into a single cluster, as represented by the
very top of the dendrogram.

The dendrogram itself does not report an optimal number of clusters; it is better
thought of as a visualization of similarity (or dissimilarity)
within a data set given a particular measure of dissimilarity. (We discuss
dissimilarity measures more thoroughly in Section~\ref{subsec:diss_meas}.). To
interpret the amount of similarity between two PK curves on
a dendrogram, it is necessary to find the vertical point where the two curves
first fuse.  It is an error to associate horizontal proximity of two curves on
the $x$-axis of the dendrogram with similarity.  For example, consider the
labeled subjects X98, X100, and Y8 on the bottom of Figure~\ref{fig:hier_ill}.
Despite the fact subjects X100 and Y8 are quite close in terms of horizontal
labels, they do not fuse until the very top of the dendrogram.  Thus,
X100 and Y8 should be considered quite dissimilar, on a relative basis, among
all PK curves within the complete data set.  On the other hand, X98 and X100
are much further apart in horizontal labels than X100 and Y8, but they fuse
much sooner vertically.  Hence, X98 and X100 should be interpreted as
relatively more similar than X100 and Y8.  Finally, horizontal
ordering of labeled subjects has no bearing on the interpretation of the
clustering outcome; it is akin to the horizontal ordering of bars on a bar
chart.  

\subsection{An Illustrative Example}
\label{subsec:hier_ill}

We demonstrate the potential
effectiveness of hierarchical clustering for PK curves  
with an illustrative example.  Consider two PK curves, $C_1$ and $C_2$,
from a one compartment linear PK model, assuming
first order absorption and first order elimination after oral administration.  Specifically, we assume
the first dosage is administered at $t=0$.  That is,
\begin{equation*}
C(t) = \frac{ D k_a }{V_c (k_a - k_{el})}
\bigg[ \exp( -k_{el} t) - \exp(-k_a t) \bigg],
\end{equation*}
where $D$ is the dosage, $V_c$ is the central volume, $k_a$ is the first order
absorption, and $k_{el} = \mathcal{K} / V_c$ (where $\mathcal{K}$ is the
clearance). The curve $C_1$ has parameters $\mathcal{K}_1 = 66.65$,
$V_{c,1} = 189.84$, and $k_{a,1} = 30.36$.  The curve $C_2$ has parameters
$\mathcal{K}_2 = 66.67$, $V_{c,2} = 93.42$, and $k_{a,2} = 43.90$.  Both curves
assume $D = 100$.  This implies $C_{\max, 1} = 0.5$ and $C_{\max, 2} = 1$ with
$\text{AUC}_1 = \text{AUC}_2 = 1.5$.

We then introduce random errors to simulate 100 realizations for each curve. 
The resulting curves are displayed in Figure~\ref{fig:hier_ill}
(top), with $C_1$ represented by the solid line and $C_2$ represented by the
dashed line.  The shaded regions around PK curves $C_1$ and $C_2$ represent
the range of 100 simulated curves. 
From the plot of the simulated PK curves in Figure 2 (top), we have two distinct PK curve groups. We assume each PK curve has 15 measurements to consider a time series data vector of length 15.
We then perform hierarchical clustering combined with Euclidean distance
to demonstrate this unsupervised
ML clustering technique is capable of identifying
the two distinct groups.  Specifically, in our hypothetical simulation
setting, we find that hierarchical clustering with Euclidean distance
obtains a mean accuracy rate of 99.9\% with a standard deviation of
0.26\% in 1{,}000 replicates. The mean Rand Index \citep{rand_1971}
was 0.998 (i.e., 1 is perfect, 0.50 indicates no contribution) with a
standard deviation of 0.0052.
An example of a resulting dendrogram appears in
Figure~\ref{fig:hier_ill} (bottom).
We will further discuss how to select the number of clusters in Section~\ref{subsec:add_cons}.

\subsection{Measures of Dissimilarity}
\label{subsec:diss_meas}

The concept of dissimilarity draws from the mathematical concept
of {\it distance} and includes some intuitive properties. For example,
any measure of distance should be nonnegative, and two identical inputs
should result in a distance of zero.
It is also preferable that a distance function be both symmetric and satisfy
the triangle inequality.  These ideas may be formalized
\citep[e.g.,][Definition 2.15]{rudin_1976}.

\cite{montero_2014} review a large set of well-established time series
dissimilarity measures.  Different distance metrics are designed to measure
different attributes, and it is important to determine a suitable
dissimilarity measure for a given data set to avoid reaching misleading conclusions.
There are four broad categories of dissimilarity measures: model-free,
model-based, complexity-based, and prediction-based \citep{montero_2014}.
Model-based approaches assume the time series data has been generated from a
time-dependent parametric model, typically an AutoRegressive Integrated Moving
Average (ARIMA).
ARIMA models assume a sequential time dependence whereby the current
realization of a time series can be explained by some number
of its past realizations.  These models often rely on assumptions, such as
{\it stationarity} and independent, identically distributed
Gaussian random errors, that are generally not
satisfied for PK curves (though the latter may be appropriate for
accounting for concentration measurement or sampling error,
an application outside the scope of our analysis).
See \citet{shumway_2006} for additional details.
Complexity-based
dissimilarity can suffer from interpretation challenges.  Finally, clustering
completed PK curves will generally not require making predictions about future
levels. We can thus consider model-free distance metrics as the best potential
candidates to support the hierarchical clustering of PK curves.

\begin{figure}[t!] 
    \centering
    \includegraphics[width=\textwidth]{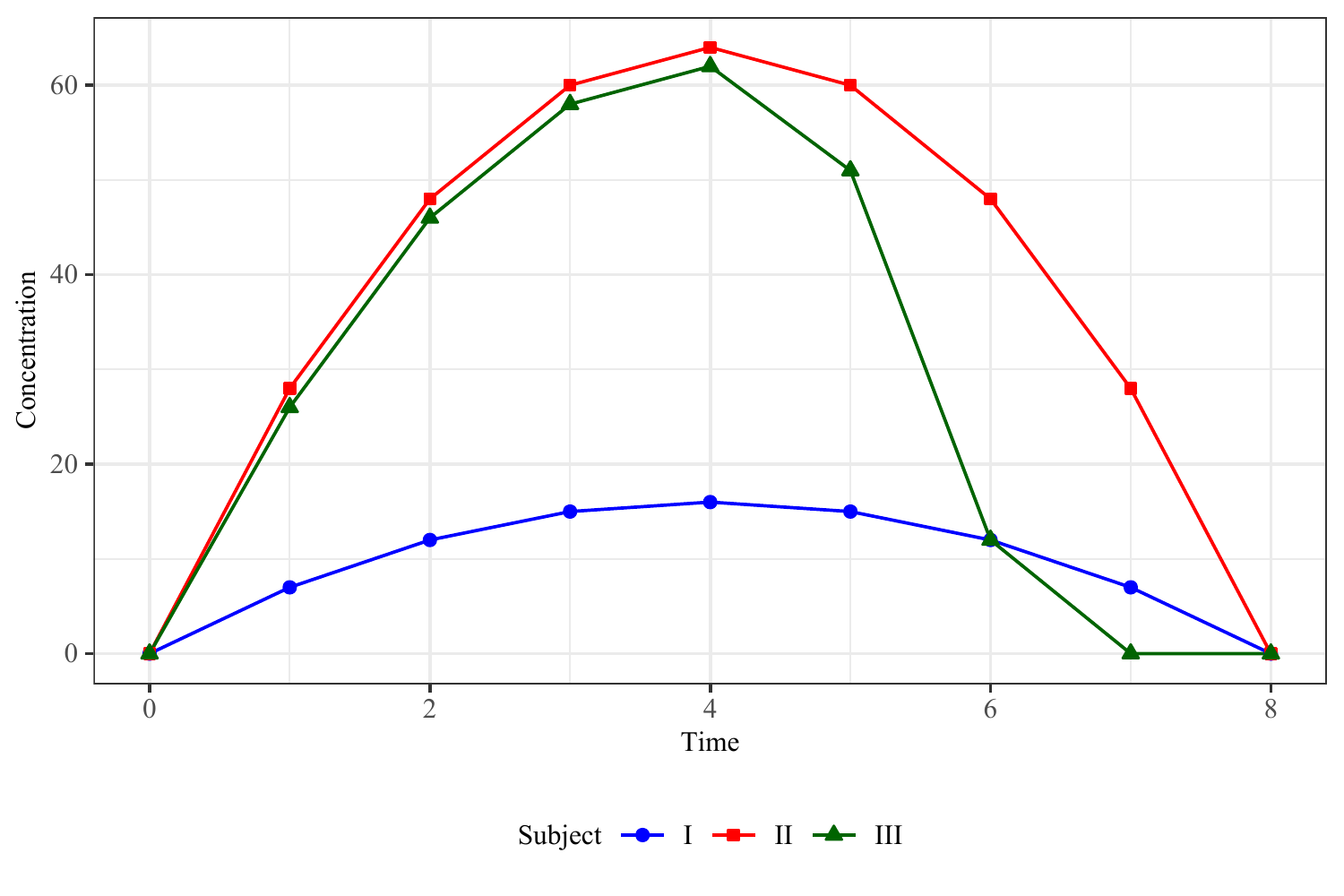}
    \caption{ {\bf Dissimilarity Measure Behavior Illustration: Shape
    vs. Structure}.
    {\scriptsize A pharmacologist comparing PK curves by AUC and $C_{\max}$
    would likely prefer Subject II and Subject III to measure as
    relatively more similar than either Subject II or Subject III with Subject
    I.
    This is the case with Euclidean, Fréchet, DTW, and CORT.
    Conversely, the correlation-based distance evaluates
    Subject I and Subject II as identical because
    Subject II’s concentration level at each time point is four times that of
    Subject I and hence linearly dependent.
    Further, Euclidean, Fréchet,
    and DTW consider Subjects II and III as most similar, followed by
    Subjects I and III, with Subjects I and II as the most dissimilar. However,
    CORT considers Subjects I and III as the most dissimilar, 
    because it considers the linear dependence between Subjects I
    and II.  Hence, it may not
    be appropriate to use a purely correlation-based measure of dissimilarity
    to cluster PK curves by geometric shape.  In another application, such as
    attempting to identify PK curves from diluted samples, however, using a
    structure-based distance (correlation, CORT) may have a more desirable
    performance. All distance metric calculations were performed using the
    \texttt{R} package \texttt{TSclust} \citep{montero_2014}.  For reference,
    we have $\{\text{AUC},C_{\max},T_{\max}\}$ of $\{84,16,4\}$,
    $\{334, 64, 4\}$, and $\{246,62,4\}$ for Subjects I, II, and III,
    respectively.  The PK parameters were calculated using the \texttt{R}
    package \texttt{PKNCA} \citep{denney_2015}.
    }
    }
    \label{fig:ill_shape}
\end{figure}

The model-free dissimilarity measures category is itself extensive
\citep{montero_2014}.  One helpful step is the further grouping of model-free
dissimilarity measures into shape-based or structure-based
\citep[e.g.,][]{lin_2009, corduas_2010}.  Shape-based dissimilarities are based
on geometric profile comparisons at a localized level. Alternatively,
structure-based dissimilarities attempt to compare underlying dependence
structures.  Figure~\ref{fig:ill_shape} presents an illustration of the
meaningful difference in results using shape- or structure-based dissimilarity
measures for PK curves.
Of the five choices we find most suitable for PK curves,
Euclidean distance, Fr\'{e}chet, and DTW are shape-based
model-free dissimilarity measures.
Correlation is a structure-based distance metric, and CORT attempts
to straddle both shape and structure. A summary of each of these five distance
metrics, including a list of pros and cons, may be found in
Table~\ref{tab:dist_summary}. For ease of exposition, we will present Euclidean
distance formally within this section and defer details of Fr\'{e}chet, DTW,
correlation, and CORT to
\hyperref[sec:diss_details]{Appendix}.  The many other model free
dissimilarity measures not suitable for PK curves fall outside our scope,
and thus further discussion has been omitted for brevity.
\citet{montero_2014} provide extended details for interested readers.

Euclidean distance is a point-by-point comparative measurement summarized into
a single number.
\textcolor{black}{Despite its canonical familiarity, we present a
formal definition for completeness and practioner-orientated readers.
Suppose $\bm{x} = \{x_1,x_2, \ldots,x_t\}^{\top}$ and
$\bm{y} = \{y_1,y_2, \ldots, y_t \}^{\top}$ are two $t$-dimensional vectors of
PK measurements, where $x_j$ and $y_j$ are measured at the same time for
$j \in \{1, \ldots, t\}$. Euclidean Distance, also sometimes called
$L_2$-distance, is defined as
\begin{equation}
d_{L_2}( \bm{x}, \bm{y}) =  \sqrt{ \sum_{i=1}^{t} (x_i-y_i )^2 }.
\label{eq:euc_dist}
\end{equation}
Because of the simplicity of \eqref{eq:euc_dist},}
Euclidean distance is
relatively straightforward to visualize and interpret as a measure of the
difference in geometric shape between two PK curves.  For example, the
Euclidean distance between $\text{Curve}_1$ and $\text{Curve}_2$ in
Figure~\ref{fig:ex_pk_dist} is simply
$d_{L_2}(\text{Curve}_1,\text{Curve}_2) =
\sqrt{ \sum_{i=1}^{t} d_i^2 }.$
Additionally, we find Euclidean distance computationally inexpensive, stable,
and capable of clustering PK curves effectively.  The major shortcoming of
Euclidean distance, however, is the data restrictions required to ensure the
point-by-point comparison is valid.
Specifically, the samples of each PK curve must be
made at the same time points or, alternatively, only measurement times
shared by all PK curves in a data set should be used.  It is not
unreasonable to suggest interpolation or extrapolation in such instances
of varied sampling times \citep[e.g.,][]{keogh_2005},
but its effectiveness for PK curves will
require further study.  At present, we avoid pursuing interpolation
techniques for fear of introducing  a difficult to measure confounding factor into our analysis,
though we acknowledge its exploration may be a viable path of future
research.

\begin{figure}[t!] 
    \centering
    \includegraphics[width=\textwidth]{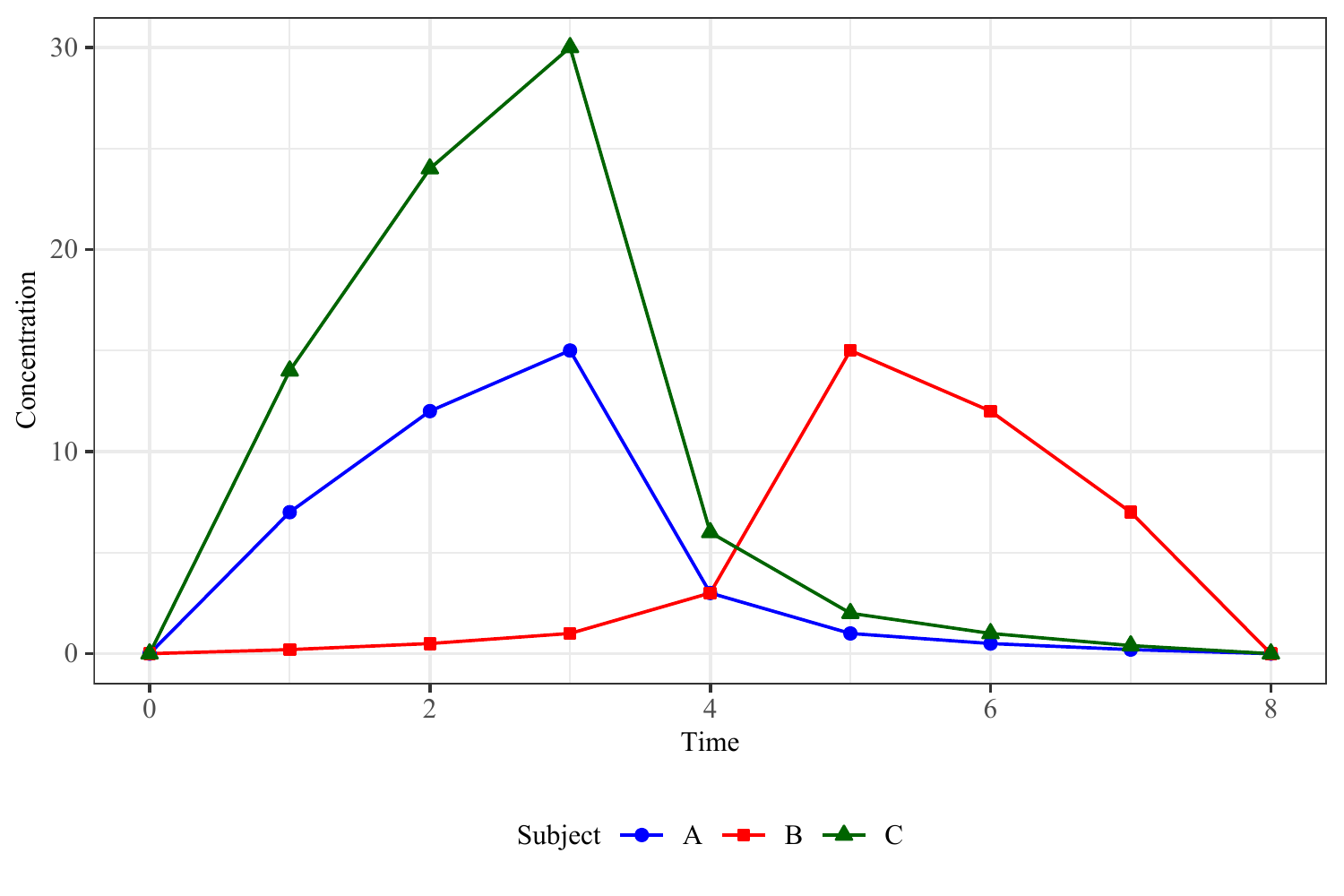}
    \caption{ {\bf Dissimilarity Measure Behavior Illustration: Fr\'{e}chet,
    DTW}.
    {\scriptsize If a pharmacologist is interested in $T_{\max}$ in addition
    to measures such as AUC and $C_{\max}$, then it may be desirable to
    classify Subjects A and B as relatively more dissimilar than Subjects A
    and C.  Fr\'{e}chet  and DTW 
    are minimization algorithms that consider more
    than single point-by-point comparisons, however, and thus they each
    identify Subjects A and B as most
    similar.  Euclidean, CORT, and the
    correlation-based distance metric each consider
    Subjects A and C as
    most similar.  Because Subject C is a twice multiple of Subject A, the
    correlation-based distance considers the two curves identical, while CORT
    also considers Subject A and C as relatively more similar than Subjects A
    and B or Subjects B and C.  All distance metric calculations were performed
    using the \texttt{R} package \texttt{TSclust} \citep{montero_2014}.
    For reference,
    we have $\{\text{AUC},C_{\max},T_{\max}\}$ of $\{37,15,3\}$,
    $\{38, 15, 5\}$, and $\{74,30,3\}$ for Subjects A, B, and C,
    respectively.  The PK parameters were calculated using the \texttt{R}
    package \texttt{PKNCA} \citep{denney_2015}.
    }
    }
    \label{fig:ill_dtw}
\end{figure}

Both DTW and Fr\'{e}chet are distance metrics that work as searching algorithms
to minimize distance over a set of comparisons between two PK curves.
\textcolor{black}{Notably, the algorithms of each allow for temporal
distortion between PK curves, which means concentration samples at different
times may end up being directly compared (see Figure~\ref{fig:dtw_app} for
an illustration). Because ``time'' is commonly one of the important factors
characterizing drug response (e.g., $C_{\max}$, $T_{\max}$), we find
this inherent property of DTW and Fr\'{e}chet to be potentially undesirable.
Indeed, such flexibility can return unexpected results in the context
of PK curves, as we demonstrate in Figure~\ref{fig:ill_dtw} with a
hypothetical example. A major distinction between
Euclidean distance and the DTW and Fr\'{e}chet distance metrics is that
the latter two will compute a valid distance between two PK curves
with either differing concentration sampling times or even a different
total number of sampling times.
Contrast this to Euclidean distance, which is restricted equivalent
to point-by-point comparisons only.  This apparent attractive property
has received some criticism \citep{ratanamahatana_2004}, and we further
find it is not enough to overlook the potential concerns identified in
Figure~\ref{fig:ill_dtw}. Finally,
because of the exhaustive searching required in the Fr\'{e}chet and DTW dissimilarity measures, they are relatively more computationally
expensive than Euclidean distance, 
especially for lengthy time series vectors within
a large observation data set.  
More details about DTW and Fr\'{e}chet are illustrated in
\hyperref[sec:diss_details]{Appendix}.}


The lone structure-based dissimilarity measure
we find applicable   to cluster PK curves is based on
Pearson's correlation coefficient \citep{golay_1998}, $d_{COR}$,
which we define formally in \eqref{eq:corr}.
The distance $d_{COR}$ considers a linear relationship between two PK curves
in evaluating dissimilarity. For example, in Figure~\ref{fig:ill_shape},
$d_{COR}$ will recognize that Subjects A and C are perfectly linearly
correlated and treat them as ``similar'', despite the notable geometric
differences. Some positive attributes of $d_{COR}$ are that it
offers a straightforward interpretation
and is computationally inexpensive.
As with Euclidean distance, it fails for two PK curves with a different number of sampling times.

Finally, the dissimilarity measure CORT, $d_{CORT}$ is both shape-based
and structure-based.  It's formal definition may be found in \eqref{eq:CORT}.
Of the five dissimilarity measures we discuss, $d_{CORT}$
is the only one that simultaneously considers both geometric shape and
structure \citep{chouakria_2007}.  
Computational time and the ability to interpret
$d_{CORT}$ can vary depending on the choice of its embedded parameters.
See the brief discussion near \eqref{eq:CORT} or \citet{chouakria_2007} for details.

To close this subsection, it is worth remarking that the four additional
model-free distance metrics
we identify as suitable for clustering PK curves
-- DTW, Fréchet, correlation, CORT --
also perform with a similar level of accuracy as that of Euclidean distance for
the 200 simulated curves
in the illustrative example of
Section~\ref{subsec:hier_ill} and Figure~\ref{fig:hier_ill}.    For more detail on
other model-free dissimilarity measures and time series dissimilarity
measures overall (including the five we discuss), see \citet{montero_2014}.
We emphasize that of all the additional
model-free dissimilarity measures in \citet{montero_2014}, we believe  that none outside of the five we summarize in Table~\ref{tab:dist_summary} are
suitable for clustering PK curves.  For reference, the
\hyperref[sec:diss_details]{Appendix} provides additional dissimilarity
measure details.

\begin{table}[tbh]
{\scriptsize
\caption{
\textbf{Summary of dissimilarity measures suitable for PK curves}.
{\scriptsize
The specific formula references
are Euclidean~\eqref{eq:euc_dist},
Fr\'{e}chet~\eqref{eq:frechet}, DTW~\eqref{eq:DTW},
correlation~\eqref{eq:corr}, and temporal~\eqref{eq:CORT}.}
Type refers to shape- versus
structure-based (see Section~\ref{subsec:diss_meas}).
The pros and cons are based on observations of the authors
in utilizing the dissimilarity measures within a ML clustering application
of PK curves.
}
	\resizebox{\textwidth}{!}{
    \centering
    \begin{tabular}{p{2cm} l c p{4cm} p{4cm}}
    	Distance & Type & Formula & Pros & Cons\\
        \toprule
        Euclidean & shape &
        $d_{L_2}(\bm{x}, \bm{y})$ &
        \begin{itemize}[nosep, left=0pt,
                before={\begin{minipage}[t]{\hsize}},
                after ={\end{minipage}}]
        \item Interpretability
        \item Computation time
        \item Stability
        \item Performance
        \end{itemize}
        &
        \begin{itemize}[nosep, left=0pt,
                before={\begin{minipage}[t]{\hsize}},
                after ={\end{minipage}}]
        \item Only valid if two PK curves share the same measurement times
        \end{itemize}
        \\
        \midrule
        Fr\'{e}chet & shape &
        $d_{F}(\bm{x}, \bm{y})$ &
        \begin{itemize}[nosep, left=0pt,
                before={\begin{minipage}[t]{\hsize}},
                after ={\end{minipage}}]
        \item Valid if two PK curves have different measurement times
        \item Valid if two PK curves have a different number of measurement
        times
        \end{itemize}
        &
        \begin{itemize}[nosep, left=0pt,
                before={\begin{minipage}[t]{\hsize}},
                after ={\end{minipage}}]
        \item Interpretability
        \item Computation time
        \item Potentially unexpected results for PK curves
        \end{itemize}\\
        \midrule
        Dynamic Time Warping (DTW) & shape &
        $d_{DTW}(\bm{x}, \bm{y})$ &
        \begin{itemize}[nosep, left=0pt,
                before={\begin{minipage}[t]{\hsize}},
                after ={\end{minipage}}]
        \item Valid if two PK curves have different measurement times
        \item Valid if two PK curves have a different number of measurement
        times
        \item Many open-source tools
        \item Many computational options
        \end{itemize}
        &
        \begin{itemize}[nosep, left=0pt,
                before={\begin{minipage}[t]{\hsize}},
                after ={\end{minipage}}]
        \item Interpretability
        \item Computation time
        \item Potentially unexpected results for PK curves
        \end{itemize}\\
        \midrule
        Correlation & structure &
        $d_{COR}(\bm{x}, \bm{y})$ &
        \begin{itemize}[nosep, left=0pt,
                before={\begin{minipage}[t]{\hsize}},
                after ={\end{minipage}}]
        \item Interpretability
        \item Computation time
        \item Identifies dependence (linear) between
        PK curves
        \end{itemize}
        &
        \begin{itemize}[nosep, left=0pt,
                before={\begin{minipage}[t]{\hsize}},
                after ={\end{minipage}}]
        \item Not suitable to cluster based on the geometric
        shape of PK curves
        \item Only valid if two PK curves have the same number of sampling times
        \end{itemize}\\
        \midrule
        Temporal & both &
        $d_{CORT}(\bm{x}, \bm{y})$ &
        \begin{itemize}[nosep, left=0pt,
                before={\begin{minipage}[t]{\hsize}},
                after ={\end{minipage}}]
        \item Only metric to consider both shape and structure of PK curves
        \end{itemize}
        &
        \begin{itemize}[nosep, left=0pt,
                before={\begin{minipage}[t]{\hsize}},
                after ={\end{minipage}}]
        \item Only valid if two PK curves have the same number of sampling times 
        \item Requires more user choice, such as $\phi_k$ or $d(\bm{x},\bm{y})$
        \end{itemize}\\
        \bottomrule
    \end{tabular}}
    \label{tab:dist_summary}
}
\end{table}

\subsection{Selecting the Number of Clusters}
\label{subsec:add_cons}

The last decision for cluster analysis is to select the appropriate number of
clusters.  
There is no current statistical consensus criteria
to arrive at a definitive validation of the resulting clusters
\citep{james_2013}. As such, deciding on the
optimal number of clusters often relies on interpretations that may vary with the nature of the underlying data or even the desired clustering resolution of
the application.

For hierarchical clustering, a visual inspection of the dendrogram is generally used (e.g., Figure~\ref{fig:hier_ill})
as a starting point.
More specifically, we decide where to ``cut'' the tree; i.e., draw a
horizontal line that intersects with the dendrogram branches.  Each
intersection between the horizontal line and the dendrogram branches indicates a cluster (i.e., all observations below the intersection point belong to that
cluster).  This is illustrated in Figure~\ref{fig:hier_ill} and discussed more
thoroughly in \citet[][Figure 10.9, pg. 392]{james_2013}.

Metrics, referred to as cluster validation indices (CVIs), are also available for estimating the quality of partitions produced by clustering algorithms and for determining the number of clusters in data. 
One such commonly used CVI is the   Calinski-Harabasz index \citep{arbelaitz_2013}, which is done by comparing the average
distance within a cluster against the average distance between clusters. The Calinski-Harabasz  is calculated by
comparing the average distance within a cluster against the average distance
between clusters.
Let $N$ be the number of
PK curves with $t$ measurement times, denoted by a vector
$\bm{x}_i = (x_1, \ldots, x_t)^{\top}$, $i = 1, \ldots, N$.  Further assume
there are $K$ clusters, each of which is denoted $c_j$, $j = 1, \ldots K$. If
we let $n_j$ denote the number of PK curves in cluster $c_j$,
$j = 1, \ldots, K$, then we can define the \textit{centroid} of a cluster as
its mean vector,
\begin{equation*}
\bar{c}_j = \frac{1}{n_j} \sum_{\bm{x}_i \in c_j} \bm{x}_i.
\end{equation*}
Similarly, the centroid of the entire PK data
set is
\begin{equation*}
\bar{\bm{x}} = \frac{1}{N} \sum_{i=1}^{N} \bm{x}_i.
\end{equation*}
The Calinski-Harabasz index for this particular set of clusters,
$C = \{c_1, \ldots, c_K\}$, denoted $\text{CH}(C)$, is then
\begin{equation}
\text{CH}(C) =
\frac{ (N-K) \sum_{c_k \in C} n_j d_{L_2}(\bar{c}_k, \bar{\bm{x}}) }
{ (K-1) \sum_{c_k \in C} \sum_{\bm{x}_i \in c_k} d_{L_2}(\bm{x}_i, \bar{c}_k) }.
\label{eq:CHC}
\end{equation}
In words, the Calinski-Harabasz CVI is a ratio of the separation between
clusters divided by the cohesion or similarity within a cluster.  A user would
then select the number of clusters that generates the highest value of
$\text{CH}(C)$.
We present an example application with PK curve data in
Section~\ref{subsec:res}.
Other CVIs and techniques are described in
Supplementary Material~\ref{subsec:sel_dets} and the references
\citep[e.g.,][]{rousseeuw_1987, kim_2005, suzuki_2006, saitta_2007,
arbelaitz_2013}.

\section{Case Study}
\label{sec:case_study}

Formal evaluations by the U.S. Food and Drug Administration (FDA) have concluded there is sufficient evidence to suggest that subgroups of patients with certain genetic variants, or genetic variant-inferred phenotypes are likely to have an altered
drug metabolism or differential therapeutic effects.  For example, the FDA recommends a reduced dosage for patients classified as a poor metabolizer (PM)
in many therapeutic areas \citep{fda_2021}.  Similar results may also be found in the literature \citep[e.g.,][]{hicks_2013, hicks_2015, lee_2018, lima_2021}.

In the original pharmacogenomic analysis of our case study data, subject
genetic information (and corresponding phenotypes) were used in conjunction
with subject PK curve samples to identify
phenotype groups that had greater exposure to the active ingredient after
drug administration (as measured by AUC).  Hence, we may perform a
validating case study analysis for the use of
hierarchical clustering with Euclidean distance on PK curve data
by performing an unsupervised ML clustering on the raw PK curve data alone
(i.e., ignoring the subject's genetic information).
In other words, if the methods
of Section~\ref{sec:methods}, using only the PK curve data, can
independently recover the same pharmacogenomic-based drug therapy
conclusion as our reference study,
then we will have demonstrable evidence of the effectiveness
of hierarchical clustering with Euclidean distance on PK curve data.
Indeed, we find not only can we validate the pharmacogenomic-based
drug therapy results with unsupervised clustering on just PK curve data,
but we also glean additional insights that would not have been easily
obtained without the use of our proposed clustering approach.  We
elaborate as follows.

\subsection{Data}
\label{subsec:data}

\begin{figure}[t!]
    \centering
    \includegraphics[width=\textwidth]{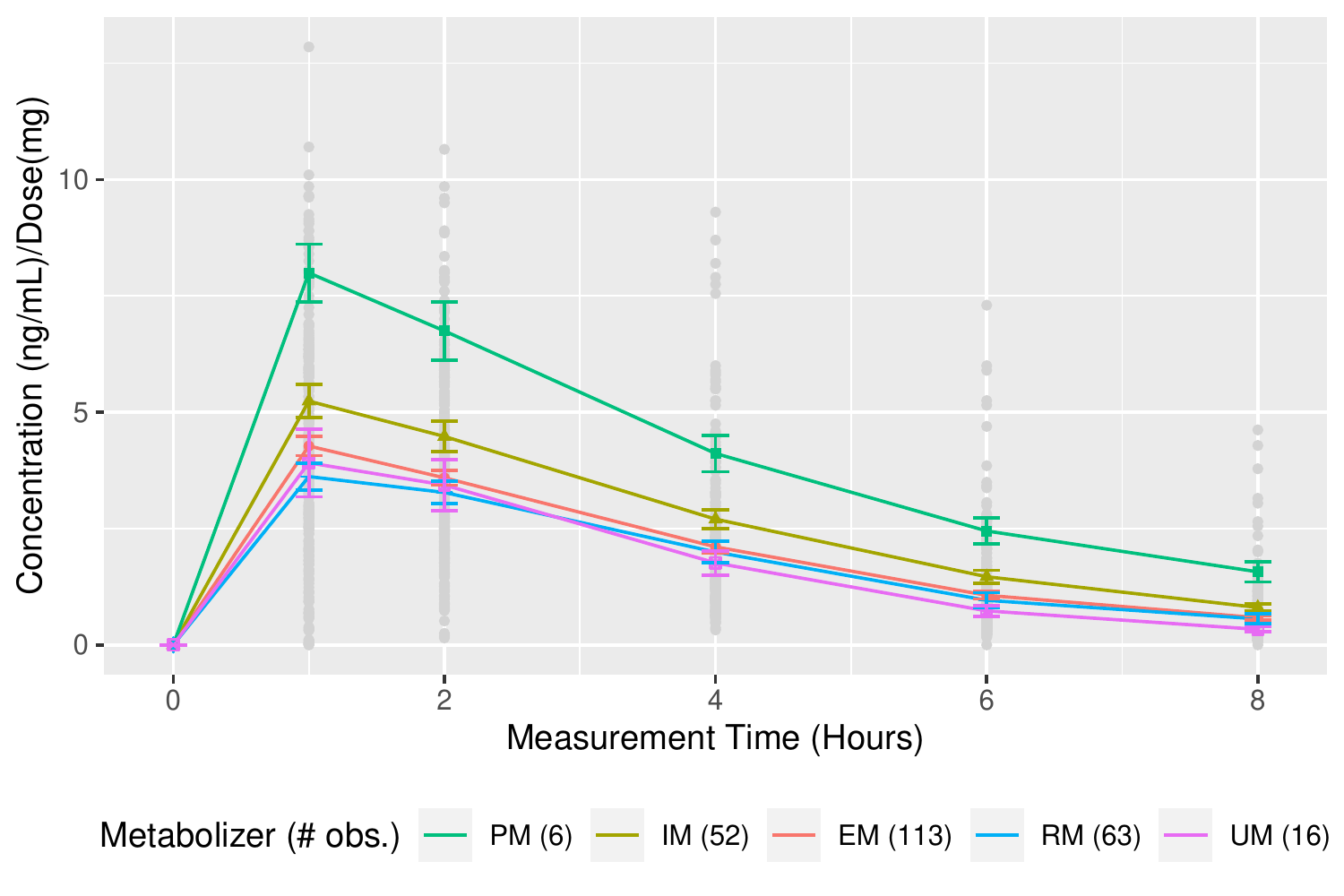}
    \caption{ {\bf Average PK Curve by  Metabolizer Classification}.
    {\scriptsize PK curves are normalized by dosage.  The CYP metabolizer
    categories are Ultra-rapid Metabolizer (UM), Rapid Metabolizer (RM),
    Extensive Metabolizer (EM), Intermediate Metabolizer (IM), and Poor
    Metabolizer (PM).  A complete PK curve is considered as one observation
    and some subjects may have more than one PK curve. The Whisker plots show one standard deviation above and below the mean of the data.
    }
    }
    \label{fig:avg_pk_metab}
\end{figure}

Our data is compiled from nine Phase 1 studies in which we have each
subject's genetic information of the cytochrome P450 (CYP) enzyme and
individual PK curve data.
At the conclusion of the nine pharmacogenomic studies,
it was suggested to use a one-half
dosage reduction for patients who are known to be 
poor metabolizers (PMs) of the CYP enzyme based
on an observed 2.3-fold increase in drug exposure in PMs compared to that in non-PMs.
The sample of 250 total observations consists of six PMs, 52 intermediate
metabolizers (IMs), 113 extensive metabolizers (EMs), 63 rapid metabolizers
(RMs), and 16 ultra-rapid metabolizers (UMs).  
Table~\ref{tab:dat_sum} summarizes basic PK curve metrics for each phenotype
and genotype.  We can see the average dose-normalized
$\text{AUC}_{\text{last}}$ is noticeably higher in the PMs in comparison
to subjects with other metabolizer classifications, and conversely, the
UMs have the lowest dose-normalized $\text{AUC}_{\text{last}}$.
The pattern of dose-normalized $C_{\max}$ is similar.
Figure~\ref{fig:avg_pk_metab} visualizes the average concentration
at each timepoint by metabolizer classification.

A complication with this data
is that the sampling times of PK concentration are not consistent for each of
the nine Phase 1 studies.  Therefore, without any adjustments
(or interpolation) to the data, only
the distance metrics Fr\'{e}chet and DTW from Table~\ref{tab:dist_summary} may
be applied.
The Fr\'{e}chet and DTW distance metrics are likely to distort time
between concentration sampling points between PK curves (e.g.,
Figures~\ref{fig:ill_dtw}, \ref{fig:dtw_app} and
\hyperref[sec:diss_details]{Appendix}),
however, and may thus produce results that
are difficult to reconcile with PK principles.  Given this and
because we desire to avoid the potential confounding effects of
interpolation, we proceed by using only the
common sampling time points across all 250 PK curve observations.  
Specifically, we find that all 250 subjects share PK measurements
at 0, 1, 2, 4, 6, and 8 hours, and we proceed using only these shared
concentration sampling time points.  For completeness, we note that
DTW and Fr\'{e}chet results on all time measurements were indeed
unsatisfactory in our testing, though the detailed results
have been omitted for brevity.

\begin{table}[t!]
\caption{
\textbf{Summary of PK Metrics by Metabolizer Status}.
{\scriptsize
The maximum concentration ($C_{\max}$) and area-under-the-time-concentration
Curve ($\text{AUC}$) are normalized by dosage.}  AUC is calculated until the
last observation ($\text{AUC}_{\text{last}}$).  A complete PK curve is
equivalent to one observation and some subjects may have more than one PK
curve.  For reference, the phenotype abbreviations are Poor Metabolizer (PM),
Intermediate Metabolizer (IM), Extensive Metabolizer (EM), Rapid Metabolizer
(RM), and Ultra-Rapid Metabolizer (UM).
}
\begin{center}
\begin{tabular}{cccccSS}
&&&
\multicolumn{2}{c}{$C_{\max}$} &
\multicolumn{2}{c}{$\text{AUC}_{\text{last}}$}\\
\multicolumn{2}{c}{CYP Enzyme} & &
\multicolumn{2}{c}{{\scriptsize(ng/ML)/Dose(mg)}} &
\multicolumn{2}{c}{{\scriptsize (ng*hr/ML)/Dose(mg)}}\\
{Phenotype} & {Genotype} & {\# Obs.} &
{Mean} & {St.Dev.} & {Mean} & {St.Dev.}\\
\hline
PM & *2/*2 & 6 & 9.03 & 1.64 & 42.09 & 8.64\\
IM & *1/*2 & 35 & 6.21 & 2.69 & 25.64 & 11.18\\
IM & *2/*17 & 17 & 6.03 & 2.17 & 26.49 & 9.39\\
EM & *1/*1 & 113 & 5.44 & 2.45 & 20.06 & 9.63\\
RM & *1/*17 & 63 & 4.88 & 2.42 & 18.49 & 12.30\\
UM & *17/*17 & 16 & 4.29 & 2.68 & 15.29 & 8.77\\
\hline
Total Sample & & \textbf{250} & \textbf{5.46} & \textbf{2.55} & \textbf{21.11}
& \textbf{11.32}\\
\hline
\end{tabular}
\end{center}
\label{tab:dat_sum}
\end{table}

\subsection{Results}
\label{subsec:res}

We find unsupervised clustering on PK curve data alone,
without using any subject genetic information,
independently validated the result of the reference study analysis.
Specifically, we find the hierarchical clustering results with Euclidean distance 
indicate that a PM subject is likely to have greater drug exposure
(as measured by cluster-level AUC) than a subject classified in the other
metabolizer categories (IM, EM, RM, UM).
This result is consistent with the proposal of the original pharmacogenomic
study which suggests  to use a one-half dosage reduction for patients of PMs of the CYP enzyme.  

\begin{figure}[t!]
    \centering
    \includegraphics[width=\textwidth]{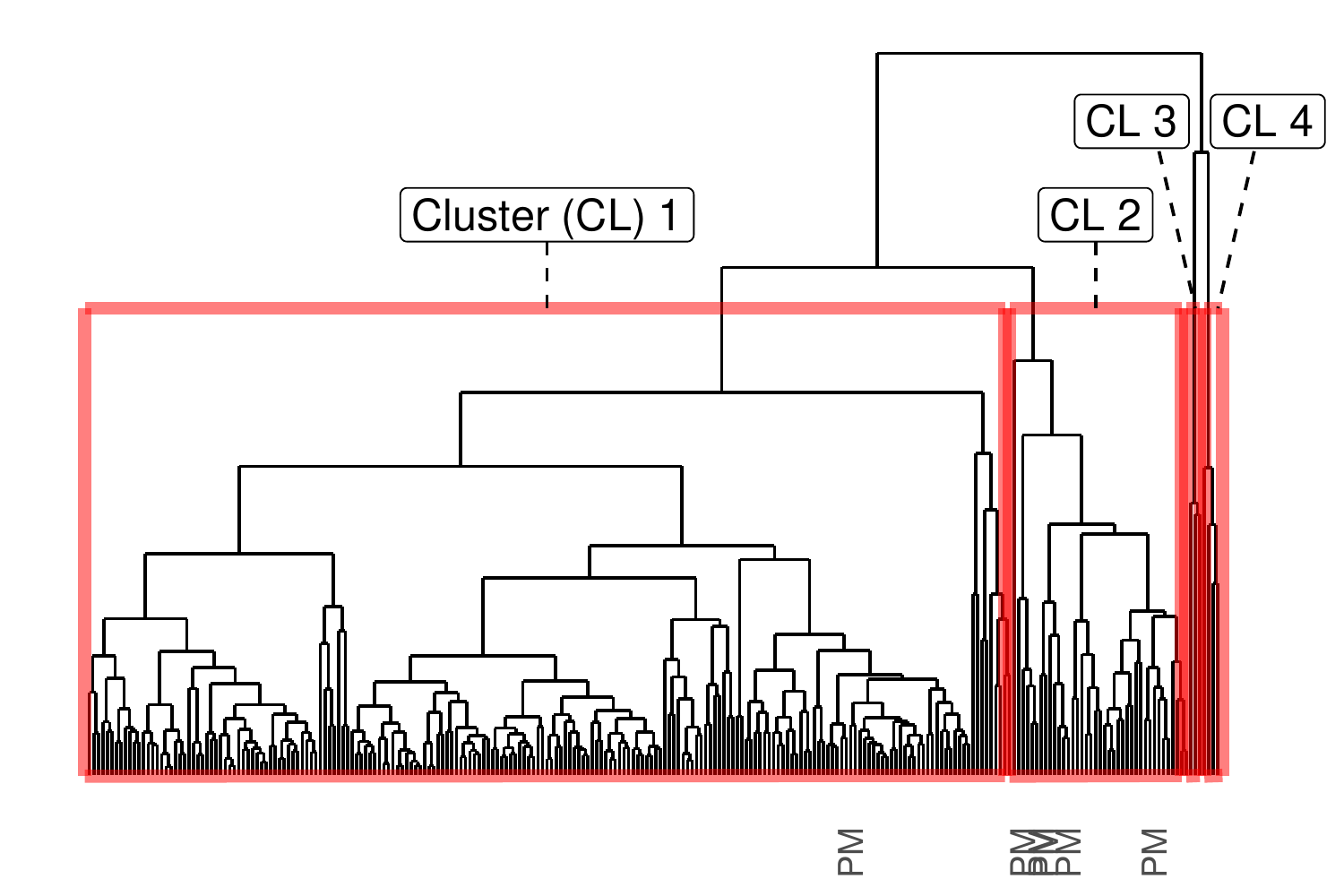}
    \caption{ {\bf Case Study Clustering Results: Dendrogram}.
    {\scriptsize The dendrogram for the 250 PK curve observations in
    Section~\ref{subsec:data} that results from clustering the distance
    matrix calculated by Euclidean distance over the shared measurement
    times.  Our recommended four clusters are highlighted by the red
    squares.  The cluster locations of the poor metabolizer (PM) subjects
    are indicated by the labels.
    All other metabolizer categories are not labeled.
    Table~\ref{tab:clust} presents summary PK metrics by cluster.
    }
    }
    \label{fig:res_dend}
\end{figure}

\begin{figure}[t!]
    \centering
    \includegraphics[scale=0.2]{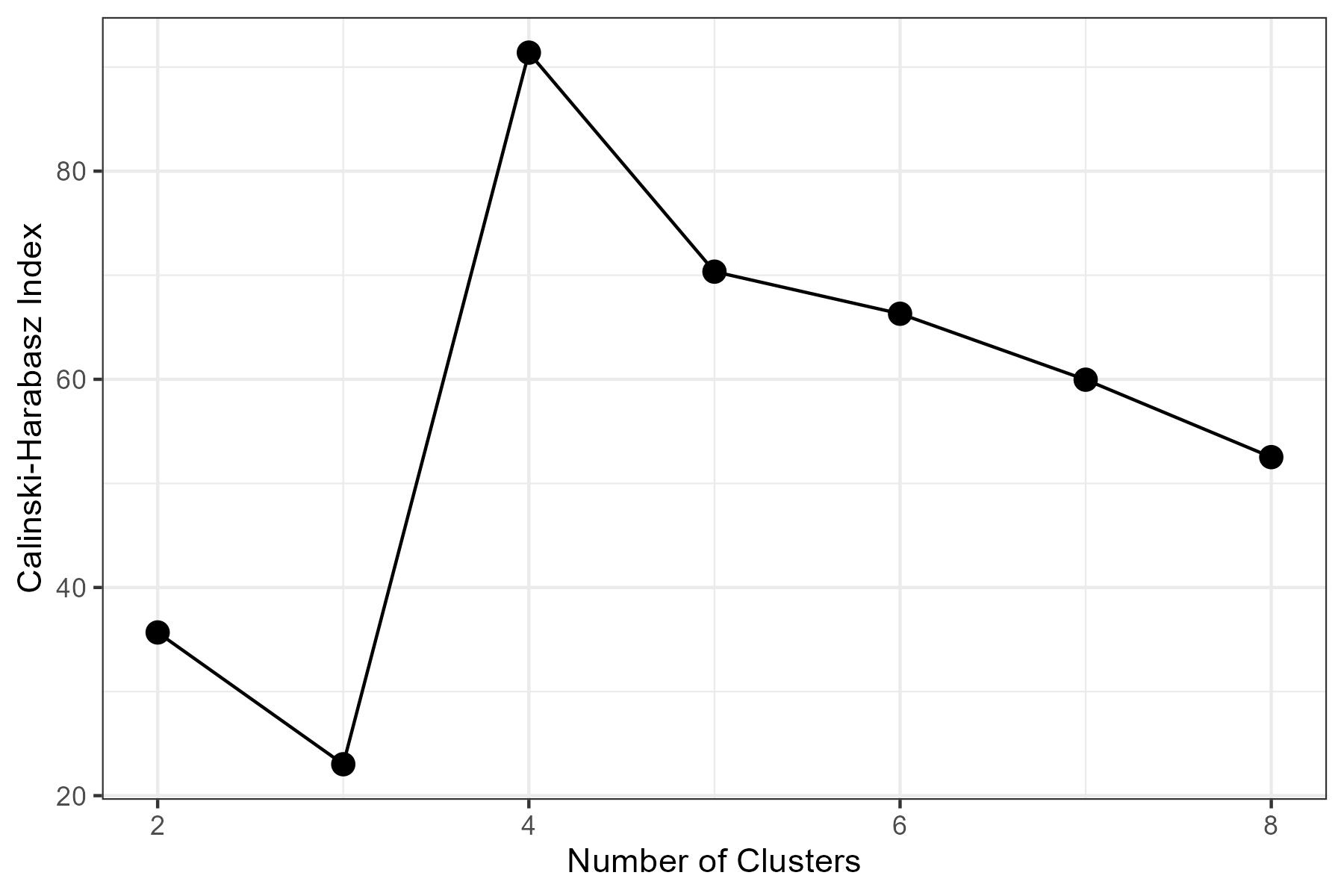}
    \caption{ {\bf Case Study Clustering Results: Calinski-Harabasz Index}.
    {\scriptsize The value of $\text{CH}(C)$ 
    \eqref{eq:CHC} for a potential number of
    clusters, $K$, for $K \in \{1, \ldots, 8\}$ based on the dendrogram
    of Figure~\ref{fig:res_dend}.  We desire to maximize
    the $\text{CH}(C)$, which implies four clusters, but subjective
    factors should also be considered (i.e., the dendrogram, see
    Figure~\ref{fig:res_dend}.)  These results were calculated with
    the \texttt{R} package \texttt{fpc} \citep{fpc}.}
    }
    \label{fig:res_ch}
\end{figure}

To obtain these results,
we employ hierarchical clustering to the data reviewed in
Section~\ref{subsec:data} by first ignoring each subject's genetic
information.  After the clusters have been established from just the PK
curve data, we then align the PK curve cluster index with each subject's
genetic information (or corresponding metabolizer status).
The optimal number of clusters is selected based on a visual
review of the dendrogram together with the Calinski-Harabasz index
($\text{CH}(C)$).
Figure~\ref{fig:res_dend}
presents a dendrogram for the sample of 250 PK curves produced by
Euclidean distance over the shared measurement times.
We also present a summary of the $\text{CH}(C)$ by number of clusters
in Figure~\ref{fig:res_ch}.  Recall that $\text{CH}(C)$ should be
maximized.  We can see that $\text{CH}(C)$ obtains a clear peak at
four clusters.  (We briefly note that it is not uncommon to observe
non-monotonic behavior of $\text{CH}(C)$ in other studies
\citep{litos_2022}.)
Visually, the dendrogram of Figure~\ref{fig:res_dend} also suggests that
four clusters would be a reasonable cutting point. Therefore, given
that $\text{CH}(C)$ is maximized at four clusters and confirmation via a
visual inspection of the vertical fuse points of the dendrogram,
we elect to use four as the optimal number of clusters.  The
four clusters are also indicated in Figure~\ref{fig:res_dend}, and
the six PMs have been labeled on the dendrogram for easy identification.
We also remark here that hierarchical
clustering provides a complete view of the structural heterogeneity of the
data via the dendrogram, which would not otherwise be easy to obtain from an
analysis based solely on summary PK metrics.

\begin{table}[t!]
\caption{
\textbf{Detailed Case Study Clustering Results}.
{\scriptsize
Summary statistics for $\text{AUC}_{\text{last}}$ and $C_{\max}$ represent the
arithmetic average by cluster with the standard deviation (St.Dev.) by cluster
provided in parenthesis. AUC is calculated until the last observation
($\text{AUC}_{\text{last}}$). Subject metabolizer status was not used to
perform the clustering; only the PK curves were used to generate the
clusters.  For reference, the abbreviations are
Poor Metabolizer (PM),
Intermediate Metabolizer (IM), Extensive Metabolizer (EM), Rapid Metabolizer
(RM), and Ultra-Rapid Metabolizer (UM).
}}
\begin{center}
{\small
\begin{tabular}{Lccccc}
Cluster & 1 & 2 & 3 & 4 & \textbf{Total}\\
\hline
\# Obs. (\%) & 204 (81.6\%) & 39 (15.6\%) & 3 (1.2\%) & 4 (1.6\%) & \textbf{250}\\
\hline
& \multicolumn{4}{c}{Distribution of Metabolizer Status by Cluster}\\
PM & 1 (16.67\%) & 5 (83.33\%) & 0 (0\%) & 0 (0\%) & \textbf{6 (100\%)}\\
IM & 35 (67.31\%) & 16 (30.77\%) & 0 (0\%) & 1 (1.92\%) & \textbf{52 (100\%)}\\
EM & 99 (87.61\%) & 11 (9.73\%) & 2 (1.77\%) & 1 (0.88\%) & \textbf{113 (100\%)}\\
RM & 55 (87.30\%) & 5 (7.94\%) & 1 (1.59\%) & 2 (3.17\%) & \textbf{63 (100\%)}\\
UM & 14 (87.50\%) & 2 (12.50\%) & 0 (0\%) & 0 (0\%) & \textbf{16 (100\%)}\\
\hline
& \multicolumn{4}{c}{Avg. (St.Dev.) PK Metrics by Cluster}\\
$\text{AUC}_{\text{last}}$ {\scriptsize (ng*hr/mL)/Dose(mg)}
& 17.12 (6.76) & 36.27 (8.28) & 60.01 (10.75) & 47.10 (10.69) &
\textbf{21.11 (11.32)}\\
$C_{\max}$ {\scriptsize (ng/mL)/Dose(mg)}
& 4.66 (1.89) & 9.04 (2.17) & 10.22 (0.45) & 7.48 (1.16) &
\textbf{5.46 (2.55)}\\
\hline
\end{tabular}
}
\end{center}
\label{tab:clust}
\end{table}

\begin{figure}[t!]
    \centering
    \includegraphics[width=\textwidth]{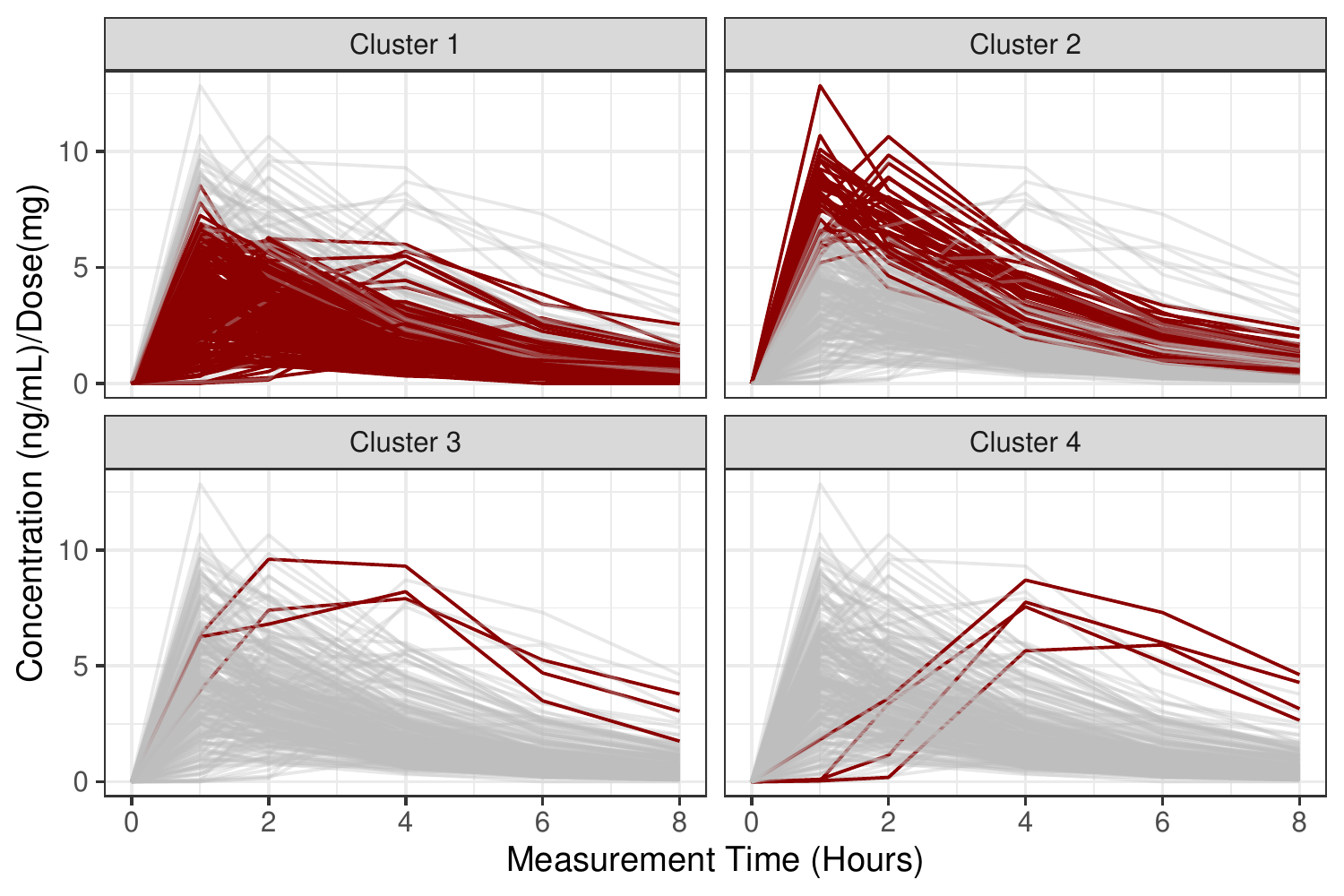}
    \caption{ {\bf Case Study Clustering Results: Visualizing Each Cluster}.
    {\scriptsize A visual representation of the four
    suggested clusters summarized in
    Figure~\ref{fig:res_dend}
    and Table~\ref{tab:clust}.
    }
    }
    \label{fig:res_vis}
\end{figure}

In our proposed four clusters summarized in Table~\ref{tab:clust},
a majority (204 out of 250) of observed PK curves
are grouped into Cluster 1.  Cluster 2 is also relatively large with a
membership of 39 observed PK curves.  Both Clusters 3 and 4 are quite small,
with 3 and 4 members, respectively.
Figure~\ref{fig:res_vis} highlights
the PK curves that comprise each cluster.
As shown in Figure~\ref{fig:res_vis}, we have performed the clustering based on the shape of the full PK curve without using any other information about the observations or subjects.
Recall our goal is to see if we can independently recover the
same potential of increased drug exposure for PMs by using unsupervised ML
on only subject PK curve data. To this end,
we connect the clustering labels with metabolizer status and
PK parameters.

The details are as follows.
Of the 204 observations grouped into Cluster 1, only 17\% (1 of 6) are PM
subjects.  Compare this to 67\% (35 of 52) of IM observations, and over
87\% for EM (99 of 113), RM (55 of 63), and UM (14 of 16) observations.
From a statistical perspective, this suggests that it is
likely that IM, EM, RM, and UM subjects would follow the PK profile of
Cluster 1, which has an average dose-normalized AUC of 17.12
(ng*hr/mL)/Dose(mg).  Most PM subjects are in Cluster 2, however
(5 of 6; 83\%).  This suggests that it is more likely that a PM subject PK
curve would follow that of Cluster 2, which has an average dose-normalized AUC
of 36.26 (ng*hr/mL)/Dose(mg).  In other words, the hierarchical clustering
analysis we performed using only PK curves  suggests that a PM subject is
likely to have a PK curve with a dose-normalized AUC of approximately double
that of observations with other phenotypes.  This corresponds very closely to
the suggestion of a dosage reduction for patients with PM in the reference studies.  
We remark that a binomial-exact hypothesis
test \citep[e.g.,][pg. 97-104]{conover_1971} 
also supports that PM subjects likely belong to Cluster 2 and
the metabolizer categories (IM, EM, RM, UM) belong to Cluster 1, the details of
which have been omitted for brevity.  

The unsupervised ML analysis also provides additional insights not reported in the reference studies. For example,
Figure~\ref{fig:res_dend} and
Table~\ref{tab:clust} indicate seven observations in Clusters 3 and 4 that have both a higher AUC and $C_{\max}$ than Cluster 2 and are not PM subjects: IM
(1), EM (3), and RM (3).  These observations may represent an important subset
of the population with a stronger reaction to the active ingredient than
the general population, or they may simply represent measurement errors or
outliers.  In either case, the granularity of hierarchical clustering suggests
seven subjects worthy of further investigation that were not identified in the reference studies.

\begin{figure}[t!]
	\centering
	\includegraphics[width=\textwidth]{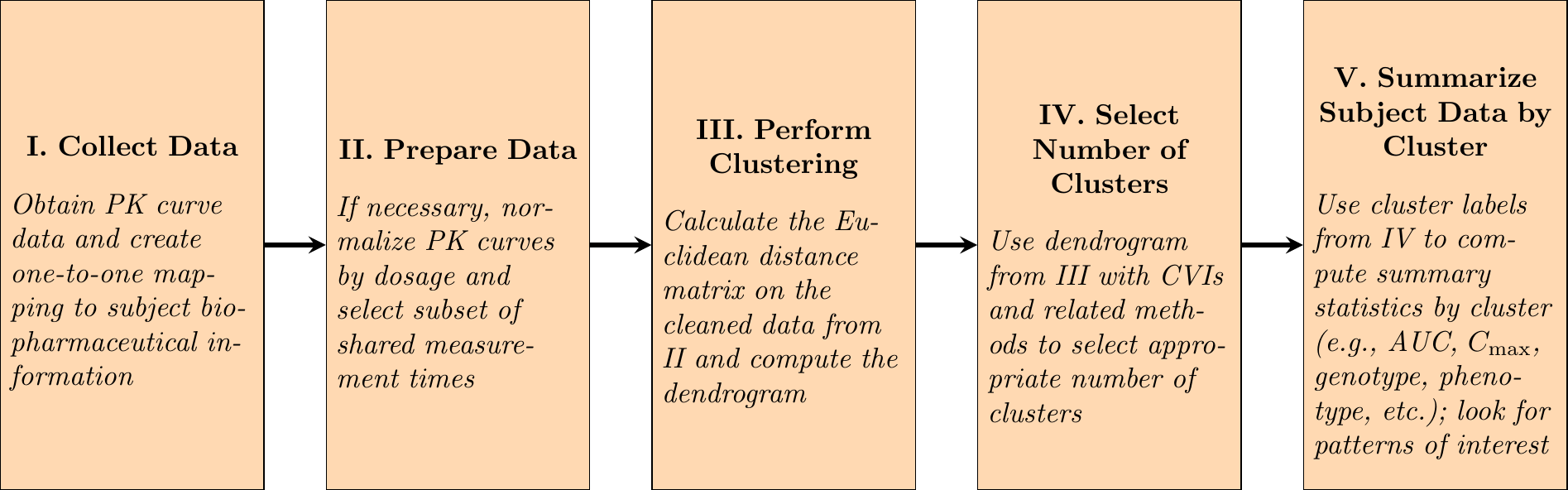}
    \caption{ {\bf Suggested PK Curve Clustering Workflow}.
    }
    \label{fig:work_flow}
\end{figure}

Furthermore, while the percentage of PM subjects in Cluster 2 is markedly higher than the percentage of PM subjects in Clusters 1, 3, or 4, there are phenotypes of each other type of metabolizer category (IM, EM, RM, UM) that are also clustered into Cluster 2, albeit in much smaller percentages versus Cluster 1.  
A further study may reveal additional factors potentially affecting absorption of the active ingredient, which may help direct the research of precision medicine into alternative directions (e.g., Do IM subjects also warrant caution or a reduced dosage in some cases?).
Although not performed here, the clustered PK curve
data could also be used to search for patterns beyond metabolizer status, such as subject weight, body mass index, age, or sex.
We find this case study reveals the clustering technique we employ can identify additional heterogeneity among subject's drug exposure. 
For the benefit of future research,
our case study process is summarized in Figure~\ref{fig:work_flow}.
As a brief comment to close the section,
we also considered the other distance metrics described in
Section~\ref{sec:methods} (DTW, Fréchet, correlation, CORT). Only DTW was able
to produce results in agreement with Euclidean distance.  For additional
detail, see the Supplemental Material.

\section{Discussion}
\label{sec:disc}

While ML and AI has attracted attention in the clinical research area in recent years, clustering techniques have not been studied or applied in understanding or interpretation of PK curves. 
The purpose of this paper was to examine the applicability and potential of clustering algorithms in the context of pharmacokinetics. We believe our paper contributes to the literature on the use of ML in drug development.
Specifically, we showed that the use of hierarchical clustering on PK curve data with
Euclidean distance over the shared concentration measurement times is a
potentially valuable enhancement to pharmacological studies.

Important outcomes of our study were twofold. 
First was to recognize that PK curve data is a time series data object 
and subsequently determine
an optimal measure of PK curve similarity.  This required narrowing
down a long list of well-established time series dissimilarity measures
\citep[e.g.,][]{montero_2014} down to five candidates: correlation,
Euclidean, Fr\'{e}chet, DTW, and CORT.  Of these five, we found that
structure-based measures (i.e., correlation and CORT) may produce
undesirable results if we are not interested in any potential linear
relationship between PK curves (e.g., Figure~\ref{fig:ill_shape}).
Further, we found the potential temporal distortion of DTW and
Fr\'{e}chet may also produce potentially misleading or difficult to
interpret results within the context of PK curves
(e.g., Figure~\ref{fig:ill_dtw}, Figure~\ref{fig:dtw_app}).
We therefore suggest Euclidean distance, which performs well,
has low computational cost, and is easy to interpret.  For convenience,
Table~\ref{tab:dist_summary} provides a complete summary of our observations
related to each of these five potential distance metrics.

Second, we present the first known case study using unsupervised learning on PK curve data.
Specifically, we utilized PK curve data from a pharmacogenomic study that
suggested a reduced dosage reduction for patients known to be PMs of
the CYP enzyme.  We found that clustering PK curve data with
hierarchical clustering based on Euclidean distance, while ignoring
subject genetic information, was able to independently arrive at the
same conclusion: PMs likely need a reduced dosage based on an
elevated drug exposure.  Additionally, our clustering results found
areas of potential further study, which were not identified in the
standard PK metric analysis reference study.

We conclude with brief comments on suggested future work.  Despite our
encouraging results, a shortcoming of our approach is that Euclidean distance
requires a one-to-one correspondence between
concentration sampling time points.  As we found,
this may require removing some concentration measurements from PK curve data,
especially if they are compiled from different clinical trial phases.
\citet{keogh_2005} suggest interpolation, but this will require further study
for PK curves.  DTW and Fr\'{e}chet allow for comparisons of time series of
different lengths, but the possible temporal distortion was not desirable in
our application.
Therefore, we suggest future research related to
finding dissimilarity measures that are suitable for PK curves with
irregular or a different number of concentration sampling time points.
At present, we leave this question open to future study,
and suggest the methods we studied herein as a current best practice.

\section{Acknowledgements}

This project was supported in part by an appointment to the Research Participation Program at the U.S.\ Food and Drug Administration by the Oak Ridge Institute for Science and Education (ORISE) through an interagency agreement between the 
U.S.\ Department of Energy and the U.S.\ Food and Drug Administration. 
Jackson Lautier’s contribution was supported in part by the
Center for Drug Evaluation and Research Intramural Funding Program and a National
Science Foundation Graduate Research Fellowship under Grant No.\ DHE 1747453. 
The simulation studies, data analyses, and manuscript drafts were completed during Jackson Lautier’s
ORISE training program at the FDA.
We further thank seminar participants at the 2022 Joint Statistical Meetings in Washington, DC for providing helpful comments.

\section{Declaration of Interest}

The present study reflects the views of the authors and should not be construed
to represent the views or recommendations of the U.S.\ Food and Drug
Administration.  The authors have no other potential conflicts of interest to
report.

{\singlespacing \footnotesize
\bibliographystyle{mcap}
\bibliography{refs}
}

\appendix  

\section*{Appendix: Measures of Dissimilarity}
\label{sec:diss_details}

We provide
additional details on the dissimilarity measures DTW, Fr\'{e}chet,
correlation, and CORT discussed in Section~\ref{subsec:diss_meas}.
Let us first consider DTW \citep{berndt_1994} in some depth,
as it is a distance metric of great popularity.
For this exposition, we will
use similar notation from \citet{giorgino_2009}, which is a particularly
helpful resource. Suppose we desire to compare two time series,
$\bm{x} = (x_1, \ldots, x_N)$ and $\bm{y} = (y_1, \ldots, y_M)$.
The counting symbol $i = 1, \ldots, N$ will only be used for indexing $\bm{x}$
and $j =1, \ldots, M$ for $\bm{y}$.
Let $d(x_i, y_j)$ be a local dissimilarity function between any pair of
elements $(x_i, y_j)$, $i = 1, \ldots, N$ and $j = 1, \ldots, M$. The DTW
algorithm rests on the \textit{warping curve}, $\pi(k)$, $k = 1, \ldots, T$.
Let $\pi_x(k) \in \{1, \ldots, N\}$ and $\pi_y(k) \in \{1, \ldots, M\}$. The
warping curve is then
\begin{equation*}
\pi(k) = \big{(} \pi_x(k), \pi_y(k) \big{)}.
\end{equation*}
The warping \textit{functions}, $\pi_x(k)$, $\pi_y(k)$, remap the indexes of
$\bm{x}$ and $\bm{y}$, respectively.  Given $\pi$, we then calculate the
average accumulated distortion between the warped time series
$\bm{x}$ and $\bm{y}$,
\begin{equation*}
d_{\pi}(\bm{x}, \bm{y}) = \sum_{k=1}^{T} d(x_{\pi_x(k)}, y_{\pi_y(k)})
m_{\pi}(k) / M_{\pi},
\end{equation*}
where $m_{\pi}(k)$ is a per-step weighting coefficient and $M_{\pi}$ is a
normalizing constant (the latter ensures the accumulated distortions are
comparable along different possible paths).  A number of constraints are imposed
on $\pi$, see \citet{giorgino_2009} for details.  Finally, we have
\begin{equation}
d_{DTW}(\bm{x}, \bm{y}) = \min_{\pi} d_{\pi}(\bm{x}, \bm{y}).
\label{eq:DTW}
\end{equation}
Informally, the goal of DTW is to pick the deformation of the time axes of
$\bm{x}$ and $\bm{y}$ which brings the two time series as close as possible to
each other.  It is interesting that the elements $x_i$, $i = 1, \ldots, N$ and
$y_j$, $j = 1, \ldots, M$ only enter into the DTW computation through
$d(x_i, y_j)$; otherwise, DTW is just an indexing algorithm.  It is typical to
choose Euclidean distance for $d(x_i, y_j)$.  Further, we do not require that
$M = N$, and so time series of different lengths and measurement times may be
used.  The results may not always align with expectations, however, especially
if a user drops the right endpoint constraint:
$\{\pi_x(T) = N, \pi_y(T) = M\}$.

\begin{figure}[t!]
	\centering
	\includegraphics[width=\textwidth]{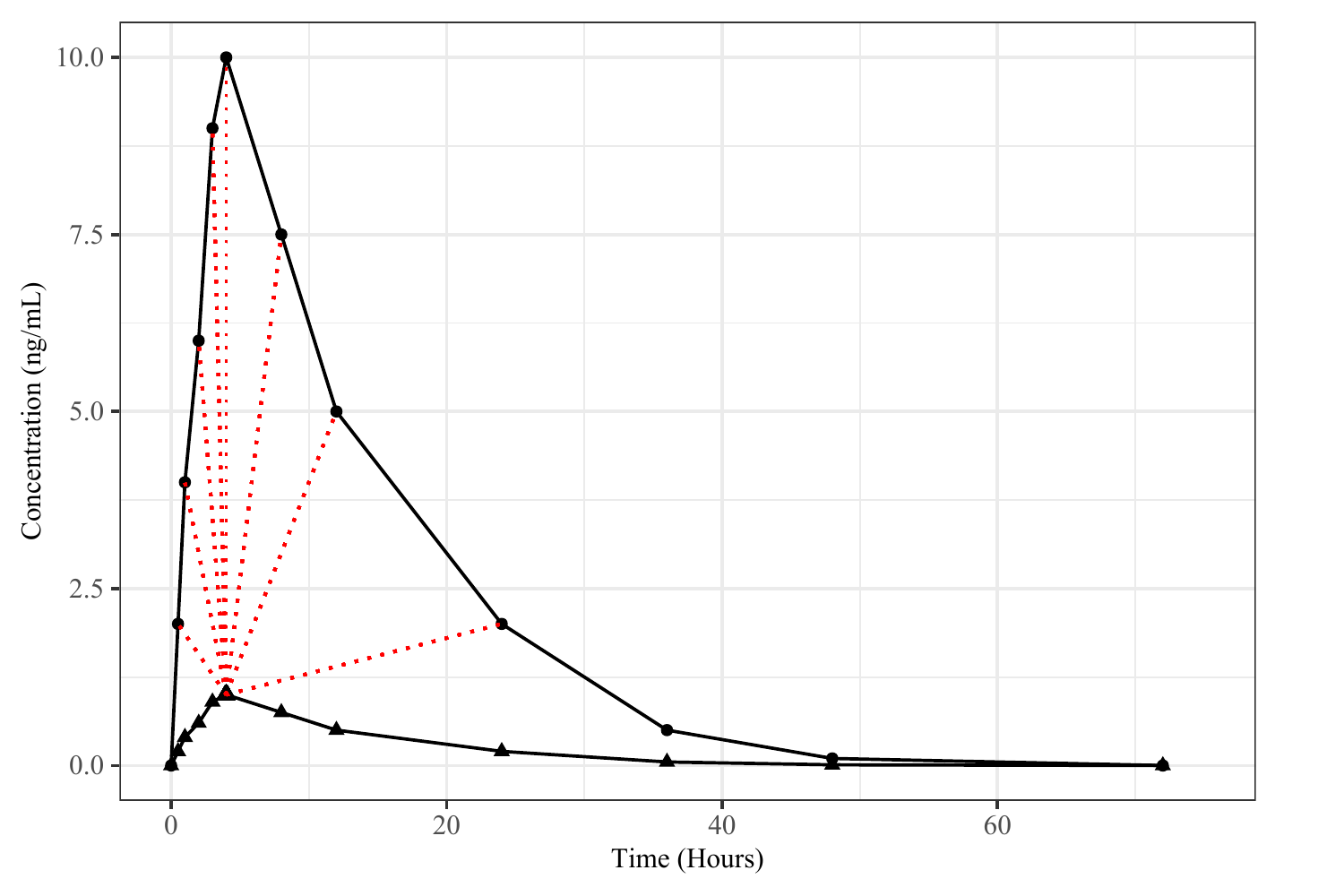}
    \caption{ {\bf Illustration of DTW Temporal Distortion}.
    {\scriptsize
    The larger PK curve (circle points) has concentration measurements,
    $\bm{x} = \{0, 2, 4, 6, 9, 10, 7.5, 5, 2, 0.5, 0.1, 0\}$, at
    hours $\{0, 0.5, 1, 2, 3, 4, 8, 12, 24, 36, 48, 72\}$.
    The smaller PK curve (triangle points) has concentration measurements
    $\bm{y} = \bm{x} / 10$ at the same measurement times.
    DTW exploits temporal distortion to minimize the distance
    between PK curves (i.e., the time index is allowed to be flexible in the
    distance calculation).  For example, it matches the maximum point on the
    smaller curve with many points on the larger curve (dashed lines). If
    $T_{\max}$ is an important factor in characterizing drug response for a
    given application, this distortion may be undesirable.
    The optimized index matching between the two curves using DTW was found
    with the default settings of the \texttt{R} package \texttt{dtw}
    \citep{giorgino_2009}.
    }
    }
    \label{fig:dtw_app}
\end{figure}

To see this, consider Figure~\ref{fig:dtw_app}, which presents
two hypothetical dosage-normalized PK curves.
The larger curve (circle points) is generated by tenfolding the smaller curve (triangle points).
In this case, statistical normalization of each PK curve would
not be appropriate (i.e., $z$-normalizing would result in two identical
time series, whereas we are interested in the relative magnitude).
This is counter to the standard advice in the literature
\citep{mueen_2016} and is a product of our specific application to
dosage-normalized PK curves.

Figure~\ref{fig:dtw_app} shows an example mapping from the DTW algorithm.
It's clear how the index remapping works to get the shortest distance.  An
analyst would need to consider if such a remapping is theoretically appropriate
for their PK curve clustering application.  Because $T_{\max}$ is an important
factor in analyzing drug response, it is possible such temporal distortion may
be undesirable.  Though not pictured, we also analyzed the DTW algorithm with
a subsample of the smaller PK curve.  The subsample has a final
measurement at hour 48, and the full sample has a final measurement at hour 72.
As such, we relaxed the end point constraint.  This resulted in the algorithm
assigning all mappings to the time zero measurement in the larger PK curve.
Once again, an analyst would need to consider if such a remapping is
theoretically appropriate for their PK curve clustering application.

We now briefly consider the Fr\'{e}chet, correlation, and
CORT dissimilarity measures.  The Fr\'{e}chet \citep{eiter_1994} dissimilarity
measure shares some general properties with DTW in that it is an index-based
searching algorithm.  Formally, let $\mathcal{Q}$ be the set of all possible
sequences of $q$ pairs that preserve the order of observations of the form
$\bm{r} = \big( (x_{a_1}, y_{b_1}), \ldots, (x_{a_q}, y_{b_q}) \big)$,
with $a_i \in \{1, \ldots, N\}$ and $b_j \in \{1, \ldots, M\}$ such that
$a_1=b_1=1$, $a_q = N$, $b_q=M$, and
$a_{i+1}=a_i$ or $a_{i+1} = a_i + 1$ and $b_{j+1}=b_j$ or $b_{j+1} = b_j+1$,
for $i \in \{1, \ldots, N-1\}$, $j \in \{1, \ldots, M-1\}$.
Then Fréchet distance is defined as
\begin{equation}
d_F( \bm{x},\bm{y}) = \min_{ \bm{r} \in \mathcal{Q} } \bigg(
\max_{i = 1, \ldots, q} \lvert x_{a_i} - y_{b_i} \rvert \bigg).
\label{eq:frechet}
\end{equation}
As with DTW, Fr\'{e}chet distance may be calculated for PK curve time series
of different lengths.

For the remainder, assuming $\bm{x}$ and $\bm{y}$ have the same length, $t$.
That is, $\bm{x} = (x_1, \ldots, x_t)$ and $\bm{y} = (y_1, \ldots, y_t)$.
The correlation dissimilarity measure is based on Pearson's correlation
coefficient \citep{golay_1998}:
\begin{equation}
d_{COR}(\bm{x}, \bm{y}) =
\sqrt{ 2 \big( 1 - \text{COR}(\bm{x}, \bm{y}) \big) },
\label{eq:corr}
\end{equation}
where
\begin{equation*}
\text{COR}(\bm{x}, \bm{y}) =
\frac{
\sum_{i=1}^{t} (x_i - \bar{x}) (y_i - \bar{y})
}{
\sqrt{ \sum_{i=1}^{t}(x_i - \bar{x})^2 }
\sqrt{ \sum_{i=1}^{t}(y_i - \bar{y})^2 }
}.
\end{equation*}
To formally define CORT, we must first
define a temporal correlation coefficient
\begin{equation}
\text{CORT}(\bm{x}, \bm{y}) =
\frac{
\sum_{i=1}^{t-1} (x_{i+1} - x_i)(y_{i+1} - y_i)
}{
\sqrt{ \sum_{i=1}^{t-1} (x_{i+1}-x_i)^2 }
\sqrt{ \sum_{i=1}^{t-1} (y_{i+1}-y_i)^2 }
}.
\label{eq:temp_cor}
\end{equation}
The ratio in \eqref{eq:temp_cor} ensures
$\text{CORT}(\bm{x}, \bm{y}) \in [-1,1]$.  The adaptive dissimilarity measure,
$d_{CORT}$, modulates between the raw values of each vector $\bm{x}$ and
$\bm{y}$ using the coefficient $\text{CORT}(\bm{x}, \bm{y})$.  Formally,
\begin{equation}
d_{CORT}(\bm{x}, \bm{y}) =
\phi_k \big( \text{CORT}(\bm{x}, \bm{y}) \big) d (\bm{x}, \bm{y}),
\label{eq:CORT}
\end{equation}
where $\phi_k$ is an adaptive tuning function dependent on parameter $k$ that
adjusts a conventional raw distance metric, $d(\bm{x}, \bm{y})$, according to
the value of $\text{CORT}(\bm{x}, \bm{y})$.  Typically, $d(\bm{x}, \bm{y})$ is
one of \eqref{eq:euc_dist}, \eqref{eq:DTW}, or \eqref{eq:frechet}.  Both
$d_{COR}$ and $d_{CORT}$ fail for two time series of different lengths.  For
additional details, we suggest \citet{montero_2014}.

\clearpage

\section*{Supplementary Materials}
\label{sec:add_detail}

\renewcommand{\thefigure}{S.\arabic{figure}}
\setcounter{figure}{0}
\renewcommand{\thetable}{S.\arabic{table}}
\setcounter{table}{0}
\renewcommand{\theequation}{S.\arabic{equation}}
\setcounter{equation}{0}
\renewcommand{\thesubsection}{S.\arabic{subsection}}
\setcounter{subsection}{0}

The following is intended as an online companion supplement to the manuscript,
{\it Clustering plasma concentration-time curves: Applications of unsupervised
learning in pharmacogenomics}.  Please attribute
any citations to the original manuscript.  This companion first presents
additional details related to Section~\ref{sec:methods}.
Specifically, Section~\ref{subsec:clust_meth} briefly reviews other clustering
techniques beyond hierarchical clustering and Section~\ref{subsec:sel_dets}
expands on techniques to select the number of
clusters.  This companion concludes by presenting additional analysis
related to the case study of Section~\ref{sec:case_study}.

\subsection{Clustering Methods}
\label{subsec:clust_meth}

For completeness, we briefly describe partition-based, model-based, and
density-based clustering techniques.

Partition-based clustering requires us to
select the number of clusters, $K$, beforehand, and each observation is then
explicitly assigned to only one cluster. Effectively, it is a combinatorial
optimization problem to minimize intra (within)-cluster distance while
maximizing inter (between)-cluster distance.  The algorithm first assigns $K$
centroids, typically by randomly selecting $K$ objects in the data set.  These
become the initial clusters, and then a distance is calculated for all objects
in the data and each centroid, after which each data object is then assigned to
its nearest centroid.  The centroids are then updated, and the process repeats
iteratively until objects no longer change clusters.  A common example is the
k-medoid algorithm along with partition-around-medoids (PAM) centroid
prototyping. Please see \citet{sarda_2019} for additional details.

For model-based clustering, a Gaussian mixture model is common, which assumes
each observation is specified through a multivariate Gaussian mixture
distribution of $G$ components.  Each component of the mixture is typically
associated with a cluster.  The task is then to fit each unknown parameter of
the underlying Gaussian distributions and $G$ through maximum likelihood
estimation.  Additional details may be found in \citet{scrucca_2016}.

Density-based clustering methods, conversely, do not assume parametric
distributions or consider variance.  Therefore, density-based clusters may be
arbitrary in shape and can handle varying amounts of noise or outliers. Like
with Gaussian mixture model-based clustering, we do not need prior knowledge of
how to determine the number of clusters.  A common technique is the DBSCAN
density-based clustering algorithm, which assigns observations to the same
cluster if they are \textit{density-reachable}.  For additional details, please
consult \citet{hahsler_2019}.

The selection process for the number of clusters for the partition-based
clustering, Gaussian Mixture Model and DBSCAN algorithms is generally like
hierarchical clustering (see Section~\ref{subsec:add_cons} and
Supplementary Material~\ref{subsec:sel_dets}). In some cases for the Gaussian Mixture Model,
however, it is common to make the selection based on the
Bayesian-Information-Criterion (BIC) \citep{schwarz_1978} in addition to
CVIs.

\subsection{Cluster Selection Criteria}
\label{subsec:sel_dets}

When selecting the number of clusters, there is often no clear choice.  The
decision can rely on interpretations that may vary with the nature of the
underlying data or even the desired clustering resolution of the practitioner
(i.e., data management versus formal research). As mentioned in
Section~\ref{sec:methods} and illustrated in Section~\ref{sec:case_study}, many
practitioners rely on a visual inspection of the dendrogram in hierarchical
clustering.

In addition, there is an automated bootstrapping method
\citep{suzuki_2006}, in which resampling procedures are employed to assess the
potential sampling uncertainty of hierarchical clusters. The main idea of the
bootstrapping approach is to obtain replicates of the dendrogram by repeatedly
applying the cluster analysis to the resampled data.  This allows a user to
identify the number of clusters that appear in a large percentage of
iterations.

Internal cluster validity indices (CVIs)  attempt to evaluate a proposed number of
clusters based only on the data used to define the clusters.  This is done by
comparing the average distance within a cluster against the average distance
between clusters.  Ideally, the average distance within a cluster will be
small, indicating similar observations have been grouped, while the average
distance between clusters will be large indicating different observations have
been treated as so.  Generally, the varying definitions of within and between
cluster distances are usually ratioed for each possible number of clusters.
This allows for a user to compare the CVI value for each possible number of
clusters.  
\cite{arbelaitz_2013} compares performance of 30 CVIs in many different environments with different characteristics. There is a selection of internal CVIs to be maximized:
Silhouette \citep{rousseeuw_1987}, 
Dunn \citep{arbelaitz_2013}, COP \citep{arbelaitz_2013}, Calinski-Harabasz
\citep{arbelaitz_2013}, Score Function \citep{saitta_2007}; or minimized:
Davies-Bouldin \citep{arbelaitz_2013},
Modified Davies-Bouldin \citep{kim_2005}.

\subsection{Case Study: Additional Details}

\begin{figure}[t!]
\caption{
\textbf{Detailed Case Study Clustering Results: Correlation Distance}.
{\scriptsize
Summary statistics for $\text{AUC}_{\text{last}}$ and $C_{\max}$ represent the
arithmetic average by cluster. AUC is calculated until the last observation
($\text{AUC}_{\text{last}}$).
The number of
clusters was selected by a visual inspection of the dendrogram in combination
with the Calinski-Harabasz Index. For reference, the abbreviations are
Poor Metabolizer (PM),
Intermediate Metabolizer (IM), Extensive Metabolizer (EM), Rapid Metabolizer
(RM), and Ultra-Rapid Metabolizer (UM).  The cluster locations of the PMs are
indicated in the dendrogram for reference.
}}
\begin{center}
{\small
\begin{tabular}{Lcccc}
\multicolumn{5}{c}{
\begin{tabular}{cc}
Dendrogram & Calinski-Harabasz (CH) Index by Cluster\\
\includegraphics[width=0.5\textwidth]{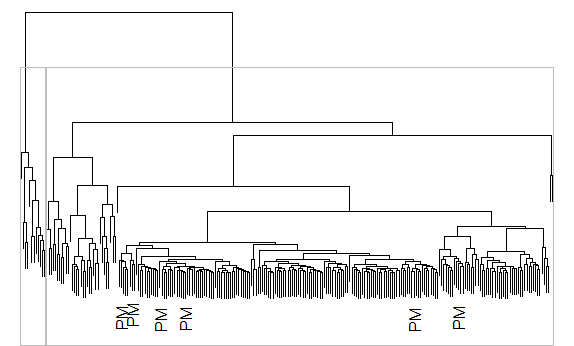}
&
\includegraphics[width=0.5\textwidth]{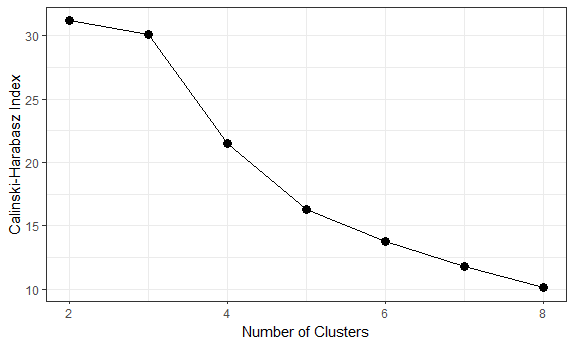}
\end{tabular}
}\\
\multicolumn{5}{c}{
\begin{tabular}{Lccc}
Cluster & 1 & 2 & \textbf{Total}\\
\hline
\# Obs. & 238 & 12 & \textbf{250}\\
\hline
& \multicolumn{2}{c}{Distribution of Metabolizer Status by Cluster}\\
PM & 6 (100\%) & 0 (0\%) & \textbf{6 (100\%)}\\
IM & 48 (92.31\%) & 4 (7.69\%) & \textbf{52 (100\%)}\\
EM & 107 (94.69\%) & 6 (5.31\%) & \textbf{113 (100\%)}\\
RM & 61 (96.83\%) & 2 (3.17\%) & \textbf{63 (100\%)}\\
UM & 16 (100\%) & 0 (0\%) & \textbf{16 (100\%)}\\
\hline
& \multicolumn{2}{c}{Avg. PK Metrics by Cluster}\\
$\text{AUC}_{\text{last}}$ {\scriptsize (ng*hr/mL)/Dose(mg)}
& 20.72 & 28.74 & \textbf{21.11}\\
$C_{\max}$ {\scriptsize (ng/mL)/Dose(mg)}
& 5.50 & 4.59 & \textbf{5.46}\\
\hline
\end{tabular}
}
\end{tabular}
}
\end{center}
\label{fig:S1}
\end{figure}

\begin{figure}[t!]
\caption{
\textbf{Detailed Case Study Clustering Results: Fr\'{e}chet Distance}.
{\scriptsize
Summary statistics for $\text{AUC}_{\text{last}}$ and $C_{\max}$ represent the
arithmetic average by cluster. AUC is calculated until the last observation
($\text{AUC}_{\text{last}}$).
The number of
clusters was selected by a visual inspection of the dendrogram in combination
with the Calinski-Harabasz Index. For reference, the abbreviations are
Poor Metabolizer (PM),
Intermediate Metabolizer (IM), Extensive Metabolizer (EM), Rapid Metabolizer
(RM), and Ultra-Rapid Metabolizer (UM).  The cluster locations of the PMs are
indicated in the dendrogram for reference.
}}
\begin{center}
{\small
\begin{tabular}{Lcccc}
\multicolumn{5}{c}{
\begin{tabular}{cc}
Dendrogram & Calinski-Harabasz (CH) Index by Cluster\\
\includegraphics[width=0.5\textwidth]{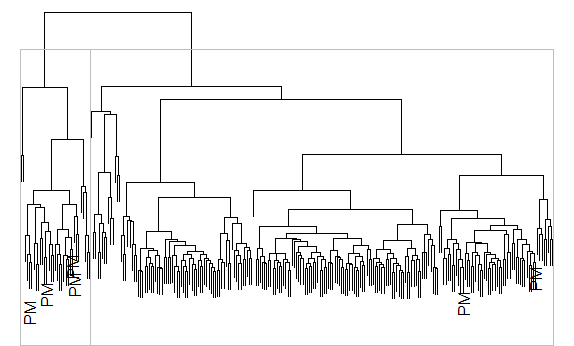}
&
\includegraphics[width=0.5\textwidth]{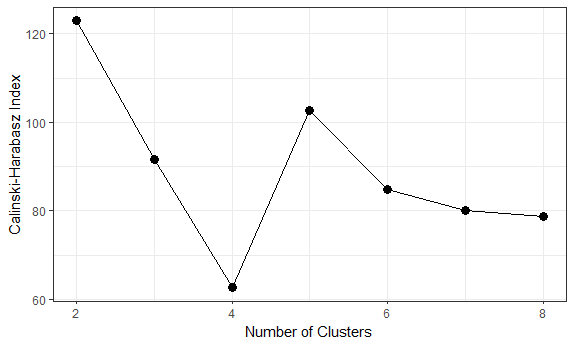}
\end{tabular}
}\\
\multicolumn{5}{c}{
\begin{tabular}{Lccc}
Cluster & 1 & 2 & \textbf{Total}\\
\hline
\# Obs. & 33 & 217 & \textbf{250}\\
\hline
& \multicolumn{2}{c}{Distribution of Metabolizer Status by Cluster}\\
PM & 4 (66.67\%) & 2 (33.33\%) & \textbf{6 (100\%)}\\
IM & 12 (23.08\%) & 40 (76.92\%) & \textbf{52 (100\%)}\\
EM & 12 (10.62\%) & 101 (89.38\%) & \textbf{113 (100\%)}\\
RM & 3 (4.76\%) & 60 (95.24\%) & \textbf{63 (100\%)}\\
UM & 2 (12.50\%) & 14 (87.50\%) & \textbf{16 (100\%)}\\
\hline
& \multicolumn{2}{c}{Avg. PK Metrics by Cluster}\\
$\text{AUC}_{\text{last}}$ {\scriptsize (ng*hr/mL)/Dose(mg)}
& 36.77 & 18.72 & \textbf{21.11}\\
$C_{\max}$ {\scriptsize (ng/mL)/Dose(mg)}
& 9.39 & 4.86 & \textbf{5.46}\\
\hline
\end{tabular}
}
\end{tabular}
}
\end{center}
\label{fig:S2}
\end{figure}

\begin{figure}[t!]
\caption{
\textbf{Detailed Case Study Clustering Results: DTW Distance}.
{\scriptsize
Summary statistics for $\text{AUC}_{\text{last}}$ and $C_{\max}$ represent the
arithmetic average by cluster. AUC is calculated until the last observation
($\text{AUC}_{\text{last}}$).  
The number of
clusters was selected by a visual inspection of the dendrogram in combination
with the Calinski-Harabasz Index. For reference, the abbreviations are
Poor Metabolizer (PM),
Intermediate Metabolizer (IM), Extensive Metabolizer (EM), Rapid Metabolizer
(RM), and Ultra-Rapid Metabolizer (UM).  The cluster locations of the PMs are
indicated in the dendrogram for reference.
}}
\begin{center}
{\small
\begin{tabular}{Lcccc}
\multicolumn{5}{c}{
\begin{tabular}{cc}
Dendrogram & Calinski-Harabasz (CH) Index by Cluster\\
\includegraphics[width=0.5\textwidth]{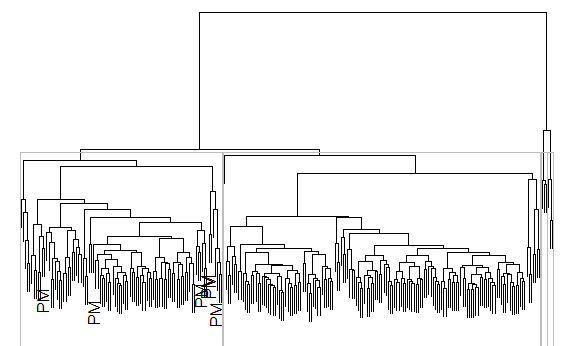}
&
\includegraphics[width=0.5\textwidth]{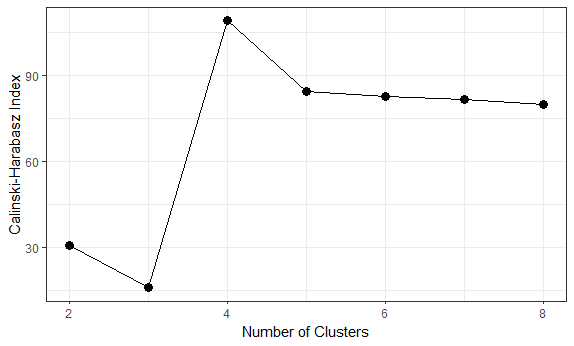}
\end{tabular}
}\\
\multicolumn{5}{c}{
\begin{tabular}{Lccccc}
Cluster & 1 & 2 & 3 & 4 & \textbf{Total}\\
\hline
\# Obs. & 95 & 149 & 3 & 3 & \textbf{250}\\
\hline
& \multicolumn{4}{c}{Distribution of Metabolizer Status by Cluster}\\
PM & 6 (100\%) & 0 (0\%) & 0 (0\%) & 0 (0\%) & \textbf{6 (100\%)}\\
IM & 32 (61.54\%) & 19 (36.54\%) & 0 (0\%) & 1 (1.92\%) & \textbf{52 (100\%)}\\
EM & 36 (31.86\%) & 75 (66.37\%) & 2 (1.77\%) & 0 (0\%) & \textbf{113 (100\%)}\\
RM & 14 (22.22\%) & 46 (73.02\%) & 1 (1.59\%) & 2 (3.17\%) & \textbf{63 (100\%)}\\
UM & 7 (43.75\%) & 9 (56.25\%) & 0 (0\%) & 0 (0\%) & \textbf{16 (100\%)}\\
\hline
& \multicolumn{4}{c}{Avg. PK Metrics by Cluster}\\
$\text{AUC}_{\text{last}}$ {\scriptsize (ng*hr/mL)/Dose(mg)}
& 28.89 & 14.97 & 55.31 & 45.24 & \textbf{21.11}\\
$C_{\max}$ {\scriptsize (ng/mL)/Dose(mg)}
& 7.68 & 3.94 & 8.80 & 6.93 & \textbf{5.46}\\
\hline
\end{tabular}
}
\end{tabular}
}
\end{center}
\label{fig:S3}
\end{figure}

\begin{sidewaysfigure}
\caption{
\textbf{Detailed Case Study Clustering Results: CORT Distance}.
{\scriptsize
Summary statistics for $\text{AUC}_{\text{last}}$ and $C_{\max}$ represent the
arithmetic average by cluster. AUC is calculated until the last observation
($\text{AUC}_{\text{last}}$).
The number of
clusters was selected by a visual inspection of the dendrogram in combination
with the Calinski-Harabasz Index. For reference, the abbreviations are
Poor Metabolizer (PM),
Intermediate Metabolizer (IM), Extensive Metabolizer (EM), Rapid Metabolizer
(RM), and Ultra-Rapid Metabolizer (UM).  The cluster locations of the PMs are
indicated in the dendrogram for reference.
}}
\begin{center}
{\small
\begin{tabular}{Lcccc}
\multicolumn{5}{c}{
\begin{tabular}{cc}
Dendrogram & Calinski-Harabasz (CH) Index by Cluster\\
\includegraphics[width=0.5\textwidth]{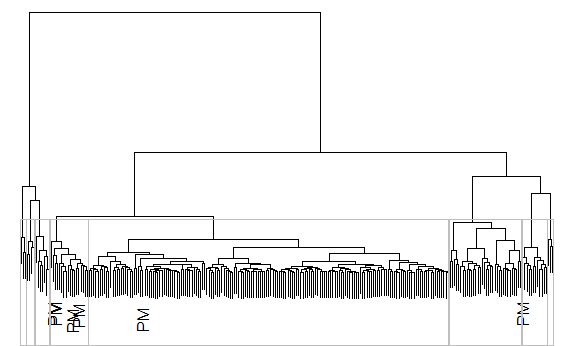}
&
\includegraphics[width=0.5\textwidth]{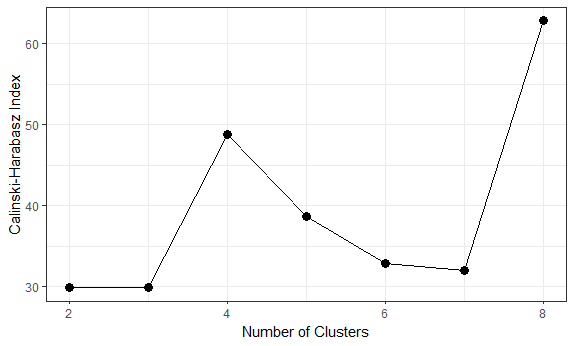}
\end{tabular}
}\\
\multicolumn{5}{c}{
{\tiny
\begin{tabular}{Lccccccccc}
Cluster & 1 & 2 & 3 & 4 & 5 & 6 & 7 & 8 & \textbf{Total}\\
\hline
\# Obs. & 169 & 34 & 18 & 12 & 3 &
7 & 4 & 3 & \textbf{250}\\
\hline
& \multicolumn{8}{c}{Distribution of Metabolizer Status by Cluster}\\
PM & 1 (16.67\%) & 0 (0\%) & 4 (66.67\%) & 1 (16.67\%) & 0 (0\%) & 0 (0\%) & 0 (0\%) & 0 (0\%) & \textbf{6 (100\%)}\\
IM & 30 (57.69\%) & 4 (7.69\%) & 5 (9.62\%) & 7 (13.46\%) & 0 (0\%) & 3 (5.77\%) & 1 (1.92\%) & 2 (3.85\%) & \textbf{52 (100\%)}\\
EM & 85 (75.22\%) & 13 (11.50\%) & 6 (5.31\%) & 1 (0.88\%) & 2 (1.77\%) & 4 (3.54\%) & 1 (0.88\%) & 1 (0.88\%) &\textbf{113 (100\%)}\\
RM & 45 (71.43\%) & 11 (17.46\%) & 1 (1.59\%) & 3 (4.76\%) & 1 (1.59\%) & 0 (0\%) & 2 (3.17\%) & 0 (0\%) & \textbf{63 (100\%)}\\
UM & 8 (50.00\%) & 6 (37.50\%) & 2 (12.50\%) & 0 (0\%)  & 0 (0\%) & 0 (0\%) & 0 (0\%) & 0 (0\%) & \textbf{16 (100\%)}\\
\hline
& \multicolumn{8}{c}{Avg. PK Metrics by Cluster}\\
$\text{AUC}_{\text{last}}$ %
& 17.55 & 16.07 & 38.12 & 39.90 & 60.01 & 18.41 & 47.10 & 33.67 & \textbf{21.11}\\
(ng*hr/mL)/Dose(mg)\\
$C_{\max}$
& 5.10 & 3.54 & 10.22 & 8.22 & 10.22 & 2.79 & 7.48 & 6.48 & \textbf{5.46}\\
(ng/mL)/Dose(mg)\\
\hline
\end{tabular}
}
}
\end{tabular}
}
\end{center}
\label{fig:S4}
\end{sidewaysfigure}

The following figures
are a summary of results related to hierarchical
clustering of a distance matrix for the PK curve data and case study of
Section~\ref{sec:case_study}.  In Section~\ref{subsec:res}, a thorough
discussion of hierarchical clustering with Euclidean distance on the shared
measurement times of the 250 PK curve observations was presented.  In this
supplement, we present numerical results for the other four measures of
dissimilarity summarized in Table~\ref{tab:dist_summary}: correlation,
Fr\'{e}chet, Dynamic time warping (DTW), and temporal (CORT). As with the
data of Section~\ref{subsec:res}, each of Figures~\ref{fig:S1},
\ref{fig:S2}, \ref{fig:S3}, and \ref{fig:S4} is a summary of clustering
analysis performed on the shared measurement times of the 250 PK curve
observations. We can see that the choice of a distance metric can have a large
influence on the results.  Of the four dissimilarity measures we consider,
only DTW performs in a way consistent with Euclidean distance.
Unless otherwise stated, the calculations were performed using the same
statistical packages referenced in the main manuscript.


\end{document}